\documentclass[runningheads]{llncs}
\usepackage[T1]{fontenc}
\usepackage{graphicx}

\usepackage{booktabs}
\usepackage{xspace}
\usepackage{hyperref}
\usepackage{xurl}
\usepackage{cleveref}
\usepackage{multirow}
\usepackage{makecell}
\usepackage[table]{xcolor} 
\usepackage{longtable}
\usepackage{subfig}
\usepackage{subcaption}
\usepackage{placeins}
\usepackage{float}
\usepackage{tabularx}
\usepackage{enumitem}
\usepackage{longtable}
\usepackage{array}
\usepackage{xspace}
\usepackage{rotating}
\usepackage{adjustbox}

\usepackage{etoolbox}
\newtoggle{morefigures}
\toggletrue{morefigures} 

\newtoggle{fullpaper}
\toggletrue{fullpaper}

\usepackage[]{todonotes}

\newcommand{\code}[1]{{\small\texttt{#1}}}

\newcommand{\rowstripegray}{gray!20}

\newcommand{\consentbase}{Base\xspace} 
\newcommand{\consentaccept}{Accept legitimate\xspace}
\newcommand{\consentdenydata}{Deny data legitimate\xspace} 
\newcommand{\consentdenyeverything}{Deny everything\xspace} 

\newcommand{\ourparagraph}[1]{\noindent\textbf{\emph{#1.}}}

\begin{document}
\title{Investigating Data Flows from Android Games}

\author{Judit Borcsiczky\orcidID{0009-0002-8430-7132} \and
Lukas Moog\thanks{Shared first authorship}\orcidID{0009-0000-5107-8260}\and
Boel Nelson\thanks{Corresponding author}\orcidID{0000-0002-9746-4885}}
\authorrunning{J. Borcsiczky et al.}
\institute{University of Copenhagen, Copenhagen, Denmark \\
\email{zwt115@alumni.ku.dk} \\ \email{smv368@alumni.ku.dk}  \\ \email{bn@di.ku.dk}}
\maketitle

\begin{abstract}
Online applications can share data with arbitrary parties, without users noticing.
In practice, these data flows could constitute privacy violations to individual users, with the lack of transparency leaving users unaware of the unwanted data flows.
To understand the spread and scale of this privacy threat, we investigate \textit{what} data is collected and \textit{where} it is sent in popular free-to-play Android games.
As part of our investigation, we conduct a measurement study where we collect and analyze over 16~GB of network traffic from 50 puzzle games.
Additionally, we explore the effectiveness of user-side countermeasures for a selection of the apps. 
We find that while most countermeasures, such as ad-blocking DNS services, decrease data flows, withholding consent and opting out of data collection can produce inconsistent behavior including increased outgoing traffic.
To encourage further analysis, we release our measurements as an open dataset. 

\keywords{Android apps \and advertisements \and data flows \and dataset \and measurement study \and privacy.}
\end{abstract}
\section{Introduction}\label{sec:introduction}

Apps and websites often collect and send personal data to other parties over the network~\cite{binns_third_2018,mayer_thirdparty_2012,vekaria_sok_2025}.
This data sharing often takes place through \textit{trackers} that, for example, collect usage data for analytics or ads, or when selling \textit{ad impressions}, for example via \textit{real-time bidding}.
While real-time bidding is intended as a way of matching advertisers with sellers of advertisement space, the auctions themselves enable parties that merely participate in an auction to collect user data~\cite{vines_exploring_2017}.
As such, real-time bidding poses a real, large-scale privacy threat where data brokers can buy and resell user data---notably, in 2024 the US Federal Trade Commission took action to ban data broker Mobilewalla from selling data from auctions as they were found to have used real-time bidding to collect and sell more than 500 million users' location data~\cite{federaltradecommission_ftc_2024}.

Modern privacy legislation, such as the General Data Protection Regulation (GDPR)~\cite{european_gdpr_2016} and the ePrivacy Directive~\cite{european_eprivacy_2002} in Europe, is designed to protect user privacy by regulating how personal data can be collected and shared.
Despite the legal protection, empirical investigations into apps show that data continues to be collected.
As an example, Reyes~et~al.'s~\cite{reyes_wont_2018} 2018 investigation found that most free children's apps potentially violate the US legislation Children's Online Privacy Protection Act (COPPA)~\cite{united_coppa_1998}, with 18.8\% of 5\,855 apps transmitting persistent identifiers to third parties.
While the changes in privacy legislation seem to have had an effect on app design, for example by the widespread adoption of cookie and consent banners following the ePrivacy Directive in Europe, the effect on privacy is unclear.
Recent work shows that data is still often collected without consent~\cite{kollnig_fait_2021}, that data is collected without \textit{explicit} consent, and that data continues to be collected after collection has been denied by the user~\cite{paci_comprehensive_2023}. 
It is also unclear if consent can be considered \textit{informed}, seeing as empirical studies observe that users perceive consent banners as containing insufficient information about what happens to their data~\cite{kyi_it_2024}.

At the core of this potential privacy problem lies the issue of transparency: if users are unaware of data collection, how can we know if the data collection is a privacy violation?
Prior studies have provided empirical evidence that many users, when made aware of trackers, perceived trackers as \textit{disturbing} or \textit{creepy}~\cite{shklovski_leakiness_2014}.
A proactive approach to avoid such reactions would be to raise awareness of the data collection, and provide user-side tools to defend against unwanted data collection.
As a step in that direction, we want to further the understanding of the data collection in today's rapidly changing ecosystem of trackers and advertisements through a measurement study, as well as a case study where we investigate the effectiveness of available countermeasures.

In this paper, we target smartphone apps and ask the question \textit{``What data is collected, and where does it go?''}, focusing specifically on apps that are financed by advertising.
In the Android ecosystem of apps, trackers are common in free-to-play games, averaging over 2.5 advertising trackers and around 2 analytics trackers per game~\cite{laperdrix_price_2022}.
As such, we focus specifically on free-to-play Android games, with the aim of creating a snapshot of the landscape in 2026.
To this end, we conduct a measurement study, collecting network traffic from the top 50 puzzle games (\Cref{sec:method}). 
We show that apps on average contact approximately 20 unique domains, with ad and analytics networks such as AppLovin being the most frequently contacted domain.
Moreover, we investigate the effectiveness of user-side defenses against data collection, where defenses either affect the usability of the apps (ad-blocking DNS servers and flight mode) or yield mixed results (refusing consent). 
Our contributions are as follows:
\begin{itemize}
    \item We publish a 16.14~GB dataset consisting of 500 network traces from triggering ads in 50 Android apps, as well as 50 separate setup runs (\Cref{sec:dataset}).
    \item We provide a snapshot of where free-to-play Android puzzle games with ads currently send data, and what data is sent (\Cref{sec:results}).
    \item We study the effectiveness of user-side countermeasures by conducting a case study of three diverse apps, including the apps with the least and most outgoing traffic, as well as one app that contains all three types of ads we target (\Cref{sec:eval_countermeasures}). Interestingly, we find that refusing consent gives mixed behavior, and can result in more network traffic generated than when only denying data collection.
    To exemplify, in one app opting out of both data and vendor preferences results in more issued requests than when simply opting out of the data preferences.
\end{itemize}

\section{Background}\label{sec:background}

\ourparagraph{Collecting data from Android apps}
Collecting data from Android apps can be done on multiple different levels.
One way to collect data is on the application layer via static analysis, for example using FlowDroid~\cite{arzt_flowdroid_2014}, or hooking into method calls using Frida~\cite{ravnaas_frida_nd}.
Another vector is to modify the runtime to allow data flows to be monitored, for example using TaintDroid~\cite{enck_taintdroid_2010,enck_taintdroid_2014}.
A less invasive strategy is to collect the network traffic, for example using Wireshark or another network protocol analyzer.
Since most approaches require the device to be rooted, and opportunities to root modern Android phones are limited~\cite{zenfyr_zenfyrdev_2026}, in this paper we choose to collect network traffic from Android emulators.
However, as most network traffic these days is encrypted, this approach still requires a way to reliably decrypt traffic.
A tool that can be used to intercept, capture and decrypt network traffic is mitmproxy~\cite{cortesi_mitmproxy_2010}, which is the tool used in this paper.

\ourparagraph{Specification for real-time bidding}
The Interactive Advertising Bureau (IAB) is an advertising trade association that develops industry standards for online advertising.
To support real-time bidding, the IAB has created an API called OpenRTB~\cite{iab_openrtb_2016}, which allows \textit{ad impressions} to be put up for bidding.
These ad impressions can have different ad types, three of which we target in this paper: \textit{banners}, \textit{rewarded} and \textit{interstitial} ads.
Banners are usually a small strip displayed at the top or bottom of the screen, rewarded ads let users opt in to watch them and are often videos, and interstitial ads are full-screen ads.
In their API, OpenRTB specifies the format of bid requests, bid responses, and enumerated lists such as content categories, ad positions and location types.

\ourparagraph{Consent banners}
Due to the EU's ePrivacy Directive and the General Data Protection Regulation (GDPR) in the European Economic Area (EEA), the UK, and Switzerland, app developers must disclose and obtain consent for the collection, sharing, and use of personal data~\cite{google_regulation_messages}.
These legal requirements also include disclosing and obtaining consent for the use of personal data for personalized advertising~\cite{google_eu_consent_policy}.
On Android, disclosures are made through pop-up messages in the app, which we refer to as \textit{consent banners}.
A user can either consent to preselected choices, or choose to manage options.
When choosing to manage options, there are two subcategories: \textit{data preferences} with a list of purposes, and \textit{vendor preferences} with a list of ad partners with whom data can be shared.
For each of the options, the app either requests \textit{consent} or discloses \textit{legitimate interest}, letting users make choices such as accept/deny consent or opt in/out of legitimate interest.
We refer to our four scenarios as \textit{\consentbase} when selecting `consent' on the main page without choosing to manage options, \textit{\consentaccept} for the case where a user does not consent but confirms legitimate interest, \textit{\consentdenydata} where the user does not consent and also opts out of the legitimate interests in the purpose list (data preferences), and \textit{\consentdenyeverything} where the user does not consent and also opts out of legitimate interest in both lists (data and vendor preferences).

\section{Method}\label{sec:method}

We conduct a measurement study to empirically investigate data flows from Android apps.
After filtering and selecting 50 apps to run on Android emulators (\Cref{sec:technical-setup}), we design and follow a protocol to collect data both in a base case (\Cref{sec:data-collection-protocol}), and when using client-side countermeasures (\Cref{sec:evaluating-countermeasures}).
The collected dataset is described in \Cref{sec:dataset}.

\subsection{Technical Setup and Scope}\label{sec:technical-setup}
We consider applications from the Google Play Store and install them on two Android emulators. 
Specifically, we use the LDPlayer emulator \cite{ldplayer} running Android 9, and an Android Studio \cite{android_studio} virtual device running Android 14. 
Both setups come with the official Google Play Store. 
Root or superuser access, which is necessary to install mitmproxy's CA certificate, is either available by default (for LDPlayer) or obtained using rootAVD \cite{root_avd} (for Android Studio).
We investigate the top 50 applications within the \emph{Puzzle} category (see full lists of included and excluded apps in
\iftoggle{fullpaper}{
Appendix~\ref{app:app-list}}{
the dataset}) in the Google Play Store that meet the following criteria:
\begin{enumerate}
    \item Ads can be triggered within five minutes of normal game play.\footnote{For some games, this requires completing the tutorial phase first.}
    \item The game is available on at least one of the emulators.
\end{enumerate}
We introduce the five-minute limit in order to keep the selection process consistent across applications and to exclude games in which advertisements cannot be reached within a practical amount of time. 
Since the ordering of games in the Google Play Store can change, we obtain the list of applications on 16/03/2026 and use this fixed list throughout the experiments.

When starting each application for the first time, we accept the terms of use and the privacy policy where required.
We also give consent for the use of personal data for personalized advertising if the option is presented. 
It is worth mentioning that not all apps present these options. 
Out of all 50 apps we investigate, 22 require accepting their terms of use or privacy policy, and 46 apps ask for consent to use personal data for personalized advertising.

\subsection{Data Collection Protocol}\label{sec:data-collection-protocol}
To make the data collection reproducible, we design a protocol to capture network traffic of an app under investigation. 
This ensures that every measurement is captured using the same systematic approach, especially regarding the stop condition described in step 2.(c). 
The protocol is as follows:

\begin{enumerate}
    \item Start the emulator and change proxy settings to use mitmproxy.
    \item For each measurement:
    \begin{enumerate}
        \item Start network traffic capture by starting mitmproxy.
        \item Start the app under investigation.
        \item Observe the first ad impression and proceed based on its type:
              \begin{enumerate}
                  \item If the ad impression is a \textit{banner}, keep the app open for 30 seconds.
                  \item If the ad impression is \textit{rewarded}, wait until the reward is granted and then close it.
                  \item If the ad impression is \textit{interstitial}, wait until the ad can be closed and close it.
              \end{enumerate}
        \item Close the app and stop traffic capture.
    \end{enumerate}
\end{enumerate}

Since each app may include different types of ad impressions, some without a trigger issued by the user (banner or interstitial ads), and some triggered by the user (rewarded ads), the ad-type-specific conditions are required. 
For rewarded ads, we close the ad after the reward has been granted, ensuring that the rewarded ad life cycle is completed. 
The documentation of several major ad providers, like Unity Ads~\cite{rewardedAd_Unity}, includes the supported length of ads, with a hard upper limit of 60 seconds. 
Others, such as Google AdMob, Meta~\cite{rewardedAd_Meta}, and AppLovin~\cite{rewardedAd_AppLovin}, specifically state that the maximum time before a rewarded ad can be skipped is 30 seconds~\cite{rewardedAd_AdMob}. 
Based on these sources and our experiences from filtering the apps, we decide to measure the banner ads' traffic for 30 seconds in order to create comparable simulations across the different ad types. 
As interstitial ads do not give rewards but eventually need to be closed to continue with the app's flow, we close them as soon as possible.

When an app is opened for the first time, its behavior may differ from everyday use, as it might configure settings and carry out other one-time activities.
We therefore separate measurements into setup runs and post-setup runs.
For each app, we measure the setup run and 10 consecutive post-setup runs for robustness. 
All measurements are done from Wi-Fi networks based in Copenhagen, Denmark.

\subsection{Evaluating Countermeasures}\label{sec:evaluating-countermeasures}
For three selected applications we evaluate countermeasures' effectiveness on privacy (i.e., limiting data collection) and usability.
We evaluate all countermeasures using the LDPlayer emulator.
The evaluated countermeasures are:

\begin{itemize}
    \item Activating flight mode.
    \item Not giving consent for the use of personal data for personalized advertising.
    \item Using an ad-blocking DNS server.
    \item Using the ad-blocking service AdGuard \cite{adguard}.
\end{itemize}

\ourparagraph{Flight mode} We enable flight mode on the emulator before starting each of the three investigated apps. 
As this countermeasure will disable network traffic, no measurements will be produced with mitmproxy.

\ourparagraph{Refusing consent}\label{sec:refusing-consent} 
In addition to the \textit{\consentbase} case where we choose \textit{accept all}, we explore withholding consent in three different stages, using \code{adb shell pm clear <package name>}, which deletes all data associated with the specified package~\cite{android_adb_commands_manual_2026}, to get a clean application state for each stage.
In each scenario we progressively disable more options to represent increasing privacy needs as follows:
\begin{description}
    \item[-- \consentaccept:] Confirm the default choices without disabling the legitimate interest options from the \textit{manage options} menu.
    \item[-- \consentdenydata:] Disable the legitimate interest options in the \textit{Data preferences} section.
    \item[-- \consentdenyeverything:] Disable all options both in the \textit{Data preferences} and \textit{Vendor preferences} sections.
\end{description}

\ourparagraph{Ad-blocking DNS server} 
DNS servers' functionality can be used to implement firewall-like behavior by returning a dummy IP address like \texttt{0.0.0.0} for all domains on a filter list. 
Public ad-blocking DNS servers make use of this strategy to provide ad blocking when configured on a user's device. 
For this countermeasure we use the public ad-blocking DNS server of AdGuard with the address \texttt{dns.adguard-dns.com}.

\ourparagraph{AdGuard} 
As a last countermeasure we install the app AdGuard~\cite{adguard} on the emulator. 
AdGuard claims to block tracking and advertising in the mobile browser, as well as in applications directly. 
We choose to rely on the measurements that AdGuard provides, as enabling mitmproxy interferes with AdGuard's approach to filtering traffic. 
We enable ad blocking, tracking protection, annoyance blocking, and DNS protection in AdGuard~\cite{adguard_2026_features}.
\section{Dataset}\label{sec:dataset}

The dataset contains regular traffic measurements for 50 applications, and countermeasure measurements for three selected applications. 
Each regular measurement contains one setup run, and ten post-setup runs. 
To study defenses, we capture traffic in four countermeasure scenarios; flight mode and AdGuard produce no capturable proxy traffic. 
In total, the dataset contains 62 measurements and 682 captured runs: regular measurements for all 50 apps plus four countermeasure measurements for each of the three selected apps.

The measurement files occupy $16.14$~GB on disk, of which $12.97$~GB belongs to the regular measurements. For these measurements, we estimate $12.07$~GB of transferred HTTPS data, corresponding to an average of $21.9$~MB per run. 
The estimate includes HTTP message bodies and metadata, but excludes lower network-layer overhead.
The dataset can be accessed and downloaded via our Harvard Dataverse repository, \textit{Measurement data for ``Investigating Data Flows from Android Games''} \cite{dataset_harvard_dataverse}.

\section{Analysis of Data Flows}\label{sec:results}

\begin{table}[tb]
\centering
\caption{Number of times domains/hosts are contacted during post-setup runs.\label{tab:domain_hostname}}
    \subfloat[Top five domains.\label{tab:domain_contacts_post_setup}]{%
    $\vcenter{\hbox{%
    {\rowcolors{2}{white}{\rowstripegray}
    \begin{tabular}{llr|}
        \toprule
        \textbf{Domain} & \textbf{Total} & \textbf{Per run} \\
        \midrule
        applovin.com & 20\,574 & 41.1 \\
        unity3d.com & 14\,910 & 29.8 \\
        doubleclick.net & 9\,796 & 19.6 \\
        inner-active.mobi & 5\,830 & 11.7 \\
        vungle.com & 5\,450 & 10.9 \\
        \bottomrule
    \end{tabular}}}}$}
\hfill
\subfloat[Top five hosts.\label{tab:hostname_contacts_post_setup}]{%
    $\vcenter{\hbox{%
    {\rowcolors{2}{white}{\rowstripegray}
    \begin{tabular}{lrr}
        \toprule
        \textbf{Host} & \textbf{Total} & \textbf{Per run} \\
        \midrule
        prod-mediate-events.applovin.com & 11\,347 & 22.7 \\
        googleads.g.doubleclick.net & 7\,234 & 14.5 \\
        ms4.applovin.com & 4\,847 & 9.7 \\
        logs.ads.vungle.com & 3\,807 & 7.6 \\
        cdn2.inner-active.mobi & 3\,691 & 7.4 \\
        \bottomrule
    \end{tabular}}}}$}
\end{table}

\begin{table}[tb]
\centering
    {
    \rowcolors{2}{white}{\rowstripegray}
    \caption{Distribution of request payload content types across all post-setup runs.}
    \label{tab:request_payload_types}
    \begin{tabularx}{0.6\textwidth}{Xr}
    \toprule
    \textbf{Payload type} & \textbf{Share (\%)} \\
    \midrule
    JSON                    & 65.1 \\
    x-www-form-urlencoded   & 15.7 \\
    Protobuf                & 8.7  \\
    Text                    & 5.1  \\
    Binary                  & 3.3  \\
    Other                   & 2.1  \\
    \bottomrule
    \end{tabularx}
    }
\end{table}

We analyze the dataset by extracting contacted domains, hosts, and endpoints, together with payload types, sizes, and contents.
Domains, hosts and endpoints are extracted from the request's \code{host} and \code{URL} fields.
Payload size is calculated from mitmproxy's \code{raw\_content} property,  containing the captured body bytes before decoding or decompression~\cite{mitm_raw_content}. 
According to RFC 9112, this is more reliable than using the \code{content-length} header, which may be absent, for example, when chunked transfer encoding is used~\cite{http_rfc9112}. 
Empty payloads are excluded from average-size calculations.

\ourparagraph{Where the data goes}
Our primary analysis considers the 500 post-setup runs, consisting of ten repeated measurements for each of the 50 applications for robustness. 
We observe $125\,979$ HTTP(S) requests to $239$ unique domains, $836$ unique hosts and $12\,582$ unique endpoints. 
These values show the diversity of the mobile advertising ecosystem, although the number of distinct companies is likely lower because companies may use multiple domains for different services.
We find that a run contains on average $219$ requests and $215.6$ responses. 
The small difference is primarily associated with certificate validation failures, and connections terminated before a response is received. 
On average, each app contacts $22.2$ unique domains, $49.4$ unique hosts, and $95.6$ unique endpoints in a run.
\Cref{tab:domain_contacts_post_setup} and~\ref{tab:hostname_contacts_post_setup} show the five most frequently contacted domains and hosts.
\iftoggle{fullpaper}{
For the full list of contacted domains, see Appendix~\ref{app:additional_results} \Cref{tab:append_full_domain_list}.
Results for setup-runs are reported separately in Appendix~\ref{app:setup-runs}.
}{

}

The listed destinations belong to major advertising and mediation providers, including  AppLovin, Unity, DoubleClick/Google, InnerActive/Digital Turbine \cite{digitalturbin_inneractive}, and Vungle. 
Although \code{unity3d.com} is the second most contacted domain, no individual Unity host appears among the five most contacted hosts, indicating that its traffic is distributed across several hosts.

\ourparagraph{Types of requests}
The payload types are obtained from the \code{content-type} header for messages with non-empty bodies.
We remove charset parameters, convert values to lowercase, and apply a mapping function that groups the normalized header values into higher-level categories such as \textit{JSON}, \textit{JavaScript}, or \textit{HTML}. 
Categories representing less than $2\%$ of messages are grouped as \textit{Other}. 
\Cref{tab:request_payload_types} shows the distribution of request payload types across all post-setup runs. 
JSON is the dominant format, followed by \textit{x-www-form-urlencoded} and the serializable format protobuf.

Most requests are encrypted, with $99.56\%$ using HTTPS, and $0.44\%$ using HTTP. 
HTTP traffic occurs in 22 applications, and the majority of these requests receive status code $503$ \textit{Service Unavailable}. 
Manual inspection did not reveal significant information in the remaining HTTP requests.
POST accounts for $56.79\%$ of requests and GET for $42.60\%$, while OPTIONS, PUT, and HEAD together make up $0.60\%$. 
Successful $2xx$ responses correspond to $94\%$ of observed responses while server errors account for $1.28\%$. 
Most remaining responses are $304$ \textit{Not Modified} responses associated with advertisement resources from \code{inner-active.mobi}.

\begin{table}[tb]
    \centering
    \caption{Distribution of search hits by category in all post-setup runs.}
    \label{tab:payload_hits_category_dist_post_setup}
    \centering
    {
    \rowcolors{2}{white}{\rowstripegray}
    \begin{tabular}{lrrr}
        \toprule
        \textbf{Category}
        & \textbf{Count}
        & \textbf{Percentage}
        & \textbf{\makecell{Hits/payload\\request}} \\
        \midrule
        Device-specific
        & 36\,694 & 33.29 & 0.61 \\
        OpenRTB specification
        & 12\,720 & 11.54 & 0.21 \\
        Personal keywords
        & 22\,339 & 20.27 & 0.37 \\
        Keywords from Caputo~et~al.~\cite{towards_user_controlled_privacy}
        & 38\,456 & 34.89 & 0.64 \\
        \bottomrule
    \end{tabular}
    }
\end{table}

\ourparagraph{What data is sent}
To identify data transmitted by applications, we inspect each request payload and decode its body according to the type specified in the \code{content-type} header. 
If structured decoding fails, we extract printable text when possible. 
We then search the decoded payloads using regular expressions covering four categories.
Based on these four categories, we create four sets of privacy-relevant keywords, with the resulting lists containing 108 keywords and regular expressions.
The complete list is provided in \iftoggle{fullpaper}{
Appendix~\ref{app:keyword_list},}{
the extended version of the paper,}
and the categories are as follows:

\begin{samepage}
\begin{itemize}
    \item Device-specific properties and values obtained through \code{adb shell getprop}, which could be used to fingerprint a device.
    \item Advertising-related terms from the OpenRTB specification, potentially indicating an ongoing ad auction.
    \item Personal information.
    \item Privacy-relevant attributes identified by prior work~\cite{towards_user_controlled_privacy}.
\end{itemize}
\end{samepage}

\begin{table}[tb]  
    \caption{Top five domains by search hits in all post-setup runs.}
    \label{tab:payload_top_domains_post_setup}
    \centering
    {
    \rowcolors{2}{white}{\rowstripegray}
    \begin{tabular}{lrr}
        \toprule
        \textbf{Domain} & \textbf{Count} & \textbf{Percentage} \\
        \midrule
        unity3d.com & 25\,774 & 23.39 \\
        amazon-adsystem.com & 11\,845 & 10.75 \\
        vungle.com & 9\,644 & 8.75 \\
        applovin.com & 8\,827 & 8.01 \\
        adjust.com & 8\,060 & 7.31 \\
        \bottomrule
    \end{tabular}
    }
\end{table}

\begin{table}[tbp]
    \centering
    {
    \rowcolors{2}{white}{\rowstripegray}
    \caption{Top and bottom five apps by search hits in all post-setup runs.}
    \label{tab:payload_top_apps_post_setup}
    \begin{tabular}{lrr}
        \toprule
        \textbf{(rank) App} & \textbf{Count} & \textbf{Percentage} \\
        \midrule
        (29) Goods Puzzle: 3D Sorting Games & 6\,015 & 5.46 \\
        (27) Crossmath® - Math Puzzle Games & 4\,637 & 4.21 \\
        (14) Arrow Out & 4\,413 & 4.00 \\
        (31) Sudoku.com - Classic Sudoku & 4\,301 & 3.90 \\
        (20) Sudoku - Classic Sudoku Puzzle & 4\,033 & 3.66 \\
        \midrule
        (38) Sheep Dash: Sheep Away & 1\,139 & 1.03 \\
        (45) Brain Puzzle 3: Crazy Mind & 1\,117 & 1.01 \\
        (13) Rope Escape Master & 1\,070 & 0.97 \\
        (26) Marble Sort! & 936 & 0.85 \\
        (61) Water SortPuz 3D: Color Sort & 695 & 0.63 \\
        \bottomrule
    \end{tabular}
    }
\end{table}

\iftoggle{morefigures}{{
\rowcolors{1}{white}{\rowstripegray}
\begin{table}[tbp]
    \caption{Top and bottom five apps by unique keywords in post-setup runs.}\label{tab:payload_top_apps_keywords_post_setup}
    \centering
    \begin{tabular}{lr}
        \toprule
        \textbf{(rank) App} & \textbf{Unique keywords} \\
        \midrule
        (35) Prison Puzzle: Brain Twist & 38 \\
        (52) Block Away - Tap Out Puzzle & 37 \\
        (24) Water Out Puzzle\texttrademark{} & 37 \\
        (59) Braindom: Brain Games Test & 36 \\
        (67) DOP 4: Draw One Part & 35 \\ \midrule
        (11) Differences - Find \& Spot It & 22 \\
        (38) Sheep Dash: Sheep Away & 22 \\
        (19) Sand Loop & 22 \\
        (43) Water Sort! Color Sort Puzzle & 20 \\
        (68) Arrow Maze - Escape Puzzle & 20 \\
        \bottomrule
    \end{tabular}
\end{table}
}}{}

\begin{sloppypar}
We search the $60\,110$ requests carrying payload using our privacy-relevant keyword set.
In total, we get $110\,209$ hits, corresponding to an average of $1.83$ matches per request with payload. 
\Cref{tab:payload_hits_category_dist_post_setup} shows the distribution across the keyword categories, indicating that most hits are device-specific or from Caputo~et~al.'s list of keywords~\cite{towards_user_controlled_privacy}.
Keyword matches mostly occur in \textit{JSON}, \textit{application/x-www-form-urlencoded}, and protobuf payloads. \iftoggle{fullpaper}{
The complete distribution is reported in \Cref{tab:payload_hits_content_type_dist_post_setup} in Appendix~\ref{app:additional_results}.
}{

}
\end{sloppypar}

\Cref{tab:payload_top_domains_post_setup} shows the top five domains by number of hits in requests across all runs. 
Three of these domains also appear among the five most frequently contacted domains in \Cref{tab:domain_contacts_post_setup}, showing partial overlap between contact frequency and the number of detected keyword matches.
\Cref{tab:payload_top_apps_post_setup} shows the apps with the most and the least hits in all post-setup runs. 
\textit{Water SortPuz 3D: Color Sort}  has the fewest matches and also the lowest average request count, at $78$ requests per post-setup run. 
This suggests that its low match count is at least partly explained by its lower traffic volume.

\Cref{tab:payload_top_apps_post_setup} ranks the applications by the total number of hits, whereas \Cref{tab:payload_top_apps_keywords_post_setup} \iftoggle{morefigures}{}{in Appendix~\ref{app:additional_results}} ranks them by the number of distinct matched keywords, where each keyword is counted at most once per application across all post-setup runs. 
We observe that the list of apps completely changes, indicating that apps with many keyword matches do not necessarily transmit a wider variety of matched attributes. 
\textit{Prison Puzzle: Brain Twist} has the highest keyword diversity, including matches such as timezone, device identifiers (model, manufacturer, operating system version, architecture), and battery information.

\section{Evaluation of Countermeasures}\label{sec:eval_countermeasures}

We select three applications for the countermeasure evaluation: \textit{Sudoku - Classic Sudoku Puzzle}, which has all three ad types (banner, interstitial and rewarded), \textit{Screw Sort Puzzle: Pin Jam 3D}, with the highest average post-setup request count ($440.9$), and \textit{Water SortPuz 3D: Color Sort}, with the lowest ($78.0$). 
We refer to them as \textit{Sudoku}, \textit{Screw Sort Puzzle}, and \textit{Water SortPuz 3D}, respectively. 
As \textit{Water SortPuz 3D} no longer functioned correctly in the Android Studio emulator, we perform countermeasure measurements in LDPlayer only, and rerecord its baseline measurements in LDPlayer to ensure comparable results.

\begin{figure}[tbp]
    \centering 
    \includegraphics[width=\linewidth,alt={A bar chart with three groups, one for Sudoku, one for Screw Sort Puzzle, and one for WaterSortPuz 3D. The bars compare the base case, the no consent/accept legitimate case, the no consent/deny data legitimate case, the no consent/deny everything case, and the DNS case.}]{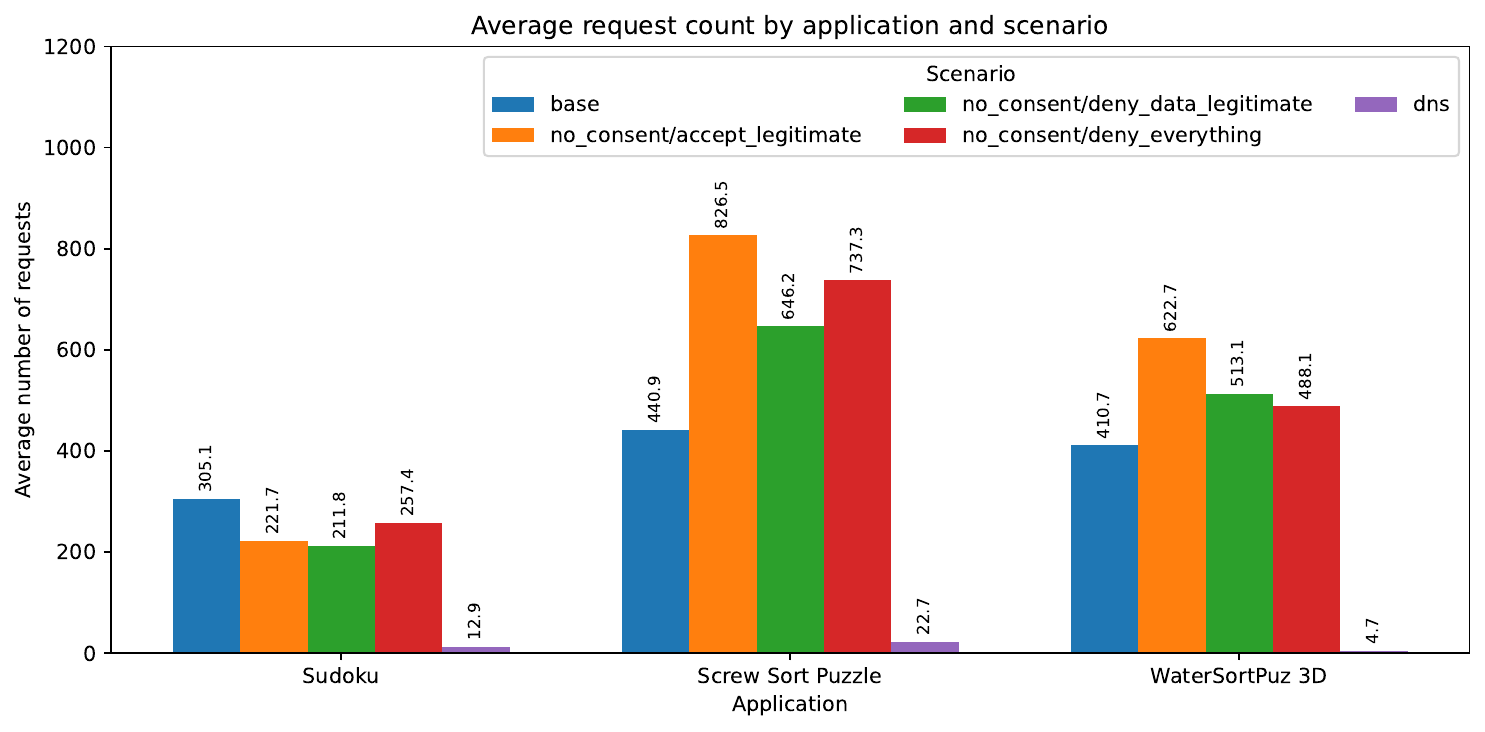}
    \caption{Comparison of the number of requests between the selected applications in the base case, when refusing consent to different degrees, and when using an ad-blocking DNS server.}
    \label{fig:avg_request_three_apps}
\end{figure}

\ourparagraph{Flight mode}
We evaluate flight mode's effectiveness by measuring network traffic and manually checking whether the selected application remains usable. 
With flight mode enabled, no traffic enters or leaves the device as expected.

All three of the selected applications remain functional. 
Flight mode prevents banner, interstitial, and rewarded advertisements from loading, which also removes access to rewards associated with advertisements. Flight mode therefore provides a complete leak-free experience for the evaluated applications, but at the cost of disabling rewarded advertisements and all other network-dependent device and application functionality.
In our pilot studies we observe that flight mode can also prevent a user from playing the game, with the app instead showing the network settings page to encourage users to turn off flight mode.

\ourparagraph{Refusing consent}
When first launched, each selected application displays a consent dialog listing the third-party vendors that can process user data. 
\textit{Sudoku} lists 297 vendors, while \textit{Screw Sort Puzzle} and \textit{Water SortPuz 3D} each list 304. 
These numbers represent potential partners rather than evidence that every listed vendor receives data during each execution, and illustrate the scale of the initial, direct flows in the advertising ecosystem exposed through the consent interface. 
Google currently permits up to 700 possible advertising partners~\cite{google_ad_partners}, but we also observe apps using a different consent dialog, for example \textit{Brick Blast - Ball Breaker} listing 1\,776 advertising partners.

We evaluate the three consent scenarios defined in \Cref{sec:refusing-consent} by manually modifying the options under \textit{Data preferences} and \textit{Vendor preferences}. 
\textit{Sudoku} presents 6 data preference and 92 vendor preference options, while \textit{Screw Sort Puzzle} presents 8 and 82, and \textit{Water SortPuz 3D} presents 8 and 80, respectively. 
Because the interface does not provide a single option for disabling all entries, users have to deselect choices individually. 
This increases the effort required to opt out or withhold consent, and is consistent with prior findings that consent interface design can influence privacy decisions~\cite{nouwens_dark_patterns_gdpr}. 
Vendors for which legitimate interest has to be disabled manually include TikTok Ad Network, Google Advertising Products, and BidMachine Inc.
The average request counts for the baseline and countermeasure scenarios are shown in \Cref{fig:avg_request_three_apps}, using the scenarios we introduced in \Cref{sec:refusing-consent}. 
We also include the data from the regular measurements with the 
\textit{\consentbase} label in the figure for comparison. \iftoggle{fullpaper}{
For a more detailed comparison, see \Cref{fig:consent_sudoku_classic_post_setup_runs},  \Cref{fig:consent_screw_sort_post_setup_runs}, and \Cref{fig:consent_watersortpuz_post_setup_runs} in Appendix~\ref{app:additional_results}. 
}{

}

We observe substantial differences between the three applications, without any clear pattern across the measurements.
For \textit{Sudoku}, every refusal scenario reduces the average request count relative to the baseline, although the strictest scenario does not produce the lowest count. 
In contrast, all refusal scenarios increase traffic for \textit{Screw Sort Puzzle} and \textit{Water SortPuz 3D}. 
Only \textit{Water SortPuz 3D} produces the expected ordering among the refusal scenarios, with fewer accepted options corresponding to fewer requests. 
These results show that stricter consent choices do not necessarily lead to proportionally lower network traffic.

The average number of domains, hosts, and endpoints is similar across the countermeasure scenarios, but this does not necessarily imply that the same destinations are contacted. 
We therefore compare the unique domains, hosts, and endpoints observed in each refusal scenario with those in the corresponding baseline to identify removed and newly introduced destinations.
\Cref{tab:consent_average_added_removed} shows the average numbers of unique domains, hosts, and endpoints that appeared only in the baseline or only in the compared consent-refusal scenario. 
Here, + denotes new values observed only in the countermeasure scenario, while - denotes removed values observed only in the baseline.

\begin{table}[tb]
    \centering
    \caption{Average numbers of removed and added domains, hosts, and endpoints for the different consent refusal scenarios.}
    \label{tab:consent_average_added_removed}
    {
    \rowcolors{2}{white}{\rowstripegray}
    \begin{tabularx}{0.8\textwidth}{lXXXXXX}
    \toprule
    & \multicolumn{2}{c}{\textbf{Domains}}
    & \multicolumn{2}{c}{\textbf{Hosts}}
    & \multicolumn{2}{c}{\textbf{Endpoints}} \\
    \cmidrule(lr){2-3}
    \cmidrule(lr){4-5}
    \cmidrule(lr){6-7}
    \textbf{Comparison}
    & \textbf{+} & \textbf{-}
    & \textbf{+} & \textbf{-}
    & \textbf{+} & \textbf{-} \\
    \midrule
    \consentaccept
    & 19.0 & 8.0
    & 61.7 & 26.0
    & 639.3 & 528.7 \\
    \consentdenydata
    & 19.0 & 6.7
    & 64.3 & 22.7
    & 646.0 & 439.3 \\
    \consentdenyeverything
    & 17.3 & 9.3
    & 60.0 & 35.0
    & 637.3 & 535.7 \\
    \bottomrule
    \end{tabularx}
    }
\end{table}

\begin{table}[tb]
\centering
\caption{Payload keyword-search hits compared to base measurements.}
\label{tab:payload_consent_summary}
    {
    \rowcolors{2}{white}{\rowstripegray}
    \begin{tabularx}{0.8\textwidth}{lXXXX}
    \toprule
    \textbf{Scenario}
    & \textbf{Payload requests}
    & \textbf{Hits}
    & \textbf{Hits/run}
    & \textbf{Hits/payload request} \\
    \midrule
    \consentbase & 6\,739 & 10\,461 & 348.70 & 1.55 \\
    \consentaccept & 11\,795 & 51\,117 & 1\,703.90 & 4.33 \\
    \consentdenydata & 9\,292 & 34\,796 & 1\,159.87 & 3.74 \\
    \consentdenyeverything & 10\,558 & 43\,690 & 1\,456.33 & 4.14 \\
    \bottomrule
    \end{tabularx}
    }
\end{table}

Across all three applications, each refusal scenario both removes destinations observed in the baseline and introduces new ones. 
This is especially visible at the endpoint level, where several hundred endpoints are added or removed on average. 
Refusing consent therefore changes not only the volume of traffic, but also the set of contacted network destinations. \iftoggle{fullpaper}{
\Cref{tab:consent_top_domains} in Appendix~\ref{app:additional_results} shows the five most frequently contacted domains in each scenario. 
}{
}
\code{applovin.com} and \code{unity3d.com} remain among the most frequently contacted domains, while \code{maticooads.com} is the most frequently contacted domain in all three scenarios. 
Although also present in the baseline measurements, \code{maticooads.com} is considerably more prominent in the consent-refusal scenarios.

To better understand whether the transmitted data becomes less privacy-sensitive, we repeat the payload analysis used for the regular measurements. 
\Cref{tab:payload_consent_summary} shows the number of searched payload requests, total keyword hits, average hits per run, and average hits per searched payload.

Disabling the legitimate-interest options under the data preferences reduces the observed keyword matches compared to accepting legitimate interest. 
However, disabling all remaining options does not reduce them further. 
Moreover, all three refusal scenarios produce more matches and a higher number of matches per payload request than the baseline. 
The relationship between stricter consent choices and the observed payload traffic is therefore not straightforward.

\iftoggle{fullpaper}{
\Cref{tab:post-setup-keyword-results-by-scenario} in Appendix~\ref{app:additional_results} provides
}{
In our extended version of the paper we provide }a category-level breakdown, reporting the number of payloads containing at least one match, the average number of unique matched keywords per affected payload, and the total number of matches.
With the exception of two OpenRTB results, the baseline produces fewer affected payloads and fewer matches than the consent-refusal scenarios across all categories.
Still, the app-level results\iftoggle{fullpaper}{
~in \Cref{tab:per-app-post-setup-keyword-results} in Appendix~\ref{app:additional_results}
}{
} show that this pattern is not consistent across applications.
\textit{Screw Sort Puzzle} accounts for most of the observed matches and shows an especially large increase in the refusal scenarios. 
\textit{Sudoku} also produces more matches in each scenario than in the baseline, although the increase is smaller. 
In contrast, \textit{Water SortPuz 3D} produces fewer affected payloads and hits under consent refusal.

The average number of hits per payload request further shows that the increases for \textit{Sudoku} and \textit{Screw Sort Puzzle} are not only caused by a larger number of payload requests. 
Both applications have higher match densities in every refusal scenario than in the baseline, whereas \textit{Water SortPuz 3D} has lower densities. 
Refusing consent therefore has no consistent effect on either the amount or density of matched payload content.

Overall, refusing consent does not consistently reduce request volume, contacted destinations, or keyword matches in the applications we evaluate. 
Our measurements and metrics suggest that refusing consent or opting out does not behave as a straightforward privacy-preserving countermeasure.
However, our measurements only show changes in the observed traffic, contacted destinations, and keyword occurrences in payloads. 
They do not show how the transmitted data is processed later, whether it is used for personalized advertising, or whether it is shared with third parties.
The broader privacy effect therefore remains unclear and appears to be application-dependent.

\ourparagraph{Ad-blocking DNS server}
\Cref{fig:avg_request_three_apps} shows that enabling the DNS server significantly decreases the average number of requests for all three applications compared to consent refusals and the \MakeLowercase{\consentbase} cases. 
Unlike flight mode, the device remains connected to the Internet and other network-dependent services remain available. 
However, no banner, interstitial, or rewarded advertisements are displayed.
Attempts to load rewarded advertisements produce errors such as \textit{video advertisement playback error} or \textit{video is not ready}, preventing users from obtaining the associated rewards. 
The contacted destinations also change relative to the baseline measurements. 
Several advertisement-related domains and hosts that appeared in the base measurements no longer appear among the most frequently contacted destinations, and the remaining destinations are contacted considerably less often.

{
\begin{table}[htb]
    \caption{Payload search results for post-setup runs grouped by scenario.}
    \label{tab:dns-post-setup-category-payload-summary}
    \centering
    \begin{tabular}{llrrr}

    \toprule
    \textbf{Scenario}
    & \textbf{Category}
    & \textbf{\makecell{Payloads\\with hit}}
    & \textbf{\makecell{Avg. unique\\keywords/payload}}
    & \textbf{Raw hits} \\
    \midrule
    
    \rowcolor{\rowstripegray}
        & Device         & 1\,809 & 1.49 & 3\,643 \\
    \rowcolor{\rowstripegray}
        & OpenRTB        & 940    & 1.10 & 1\,162 \\
    \rowcolor{\rowstripegray}
        & Caputo et al.\cite{towards_user_controlled_privacy} & 1\,938 & 1.52 & 3\,625 \\
    \rowcolor{\rowstripegray}
    \multirow{-4}{*}{Base}
        & Personal       & 1\,293 & 1.43 & 2\,031 \\

        & Device         & 94 & 1.49 & 156 \\
        & OpenRTB        & 30 & 1.00 & 30  \\
        & Caputo et al.\cite{towards_user_controlled_privacy} & 56 & 2.32 & 134 \\
    \multirow{-4}{*}{DNS}
        & Personal       & 44 & 1.00 & 44  \\
    \bottomrule
    \end{tabular}
\end{table}
} \Cref{tab:dns-post-setup-category-payload-summary} presents the payload results by keyword category.
DNS blocking substantially reduces the number of affected payloads and total keyword matches across all categories. 
Although the number of hits per searched payload is similar, the overall amount of outgoing payload data containing the specified keywords is much lower in the DNS scenario.
\Cref{tab:per-app-dns-post-setup-keyword-results} presents the app-level results, comparing the DNS scenario with the base case.

\begin{table}[tb]
\centering
\caption{Privacy-relevant keywords for DNS blocking compared to base case.}
\label{tab:per-app-dns-post-setup-keyword-results}
    \begin{tabularx}{0.95\textwidth}{lXXXXX}
    \toprule
    \textbf{App} & \textbf{Scenario} & \textbf{\makecell[l]{Payload\\requests}} & \textbf{\makecell[l]{Requests\\with hits}} & \textbf{\makecell[l]{Raw\\hits}} & \textbf{\makecell[l]{Hits/\\payload\\request}} \\
    \midrule
    \rowcolor{\rowstripegray}
        & Base & 1\,669 & 607 & 4\,033 & 2.42 \\
    \rowcolor{\rowstripegray}
    \multirow{-2}{*}{Sudoku}
        & DNS & 59 & 58 & 187 & 3.17 \\
    \midrule
    \rowcolor{white}
        & Base & 2\,618 & 560 & 2\,541 & 0.97 \\
    \rowcolor{white}
    \multirow{-2}{*}{\makecell[l]{Screw Sort Puzzle}}
        & DNS & 207 & 24 & 88 & 0.43 \\
    \midrule
    \rowcolor{\rowstripegray}
        & Base & 2\,452 & 907 & 3\,887 & 1.59 \\
    \rowcolor{\rowstripegray}
    \multirow{-2}{*}{\makecell[l]{Water\\SortPuz 3D}}
        & DNS & 47 & 26 & 89 & 1.89 \\
    \bottomrule
    \end{tabularx}
\end{table}

Across all three applications, DNS blocking substantially reduces both the number of payload requests and the total number of keyword matches. 
However, the number of matches per payload request does not decrease uniformly: it drops for \textit{Screw Sort Puzzle}, but increases for \textit{Sudoku} and \textit{Water SortPuz 3D}. 
As such, DNS blocking reduces the overall traffic volume, but shows inconsistent results when it comes to reducing the amount of privacy-relevant data transferred.
Overall, the DNS-based blocking effectively reduces outgoing payload requests and observed keyword matches. 
Still, this effectiveness comes at the cost of limited user control: when using a public ad-blocking DNS server the provider determines the filtering rules, and users may have limited ability to whitelist individual destinations.
In practice, this makes the countermeasure largely all-or-nothing.

\ourparagraph{AdGuard}
Across all setup and post-setup runs, AdGuard blocks 17\,900 ads, 2\,600 trackers, and reports 21\,800 requests in total. \Cref{tab:adguard_top_companies} shows the number of blocked ads for the top five companies. 
For the first four entries, every reported request is classified as an advertisement by AdGuard, whereas \code{dailyinnovation.biz} is partly classified as advertising (624 of 648 requests).

{
\rowcolors{2}{white}{\rowstripegray}
\begin{table}[tb]
\centering
\caption{Top five advertising companies blocked by AdGuard.}
\label{tab:adguard_top_companies}
    \begin{tabular}{lrr}
    \toprule
    \textbf{Company} & \textbf{Blocked ads} & \textbf{Per run} \\
    \midrule
    AppLovin & 8\,700 & 263.6 \\
    appsflyersdk.com & 4\,600 & 139.4 \\
    DoubleClick & 1\,100 & 33.3 \\
    Amazon Advertising & 716 & 21.7 \\
    dailyinnovation.biz & 624 & 18.9 \\
    \bottomrule
    \end{tabular}
\end{table}
}

A cautious comparison can be made for AppLovin, since it appears as the most frequent entry in both the base domain analysis and the AdGuard company statistics. 
In the base measurements, the average number of requests to \code{applovin.com} is $42.2$ during the setup runs and $41.1$ during the post-setup runs. 
In the AdGuard measurements, AppLovin accounts for $263.6$ blocked ad requests per run on average. 
These values are not directly equivalent, since the base analysis is grouped by domain while AdGuard groups traffic by company. 
However, the difference still suggests that blocking may lead to repeated ad-loading attempts. 
One possible explanation is waterfall mediation, a technique commonly used in mobile advertising, in which ad networks are contacted sequentially until an impression is filled~\cite{unity_2025_waterfall}. 
If an ad request is blocked, the SDK may continue attempting to load an advertisement, still using this sequential approach, which could explain the increase in requests.
AdGuard further estimates that blocking these requests saved approximately 2.4~GB of data during the measurements.

From a usability perspective, AdGuard may provide users more flexibility.
Unlike the ad-blocking DNS server countermeasure, AdGuard allows users to whitelist selected destinations. 
This may enable users to preserve specific functionality, such as rewarded advertisements, while continuing to block other advertising or tracking traffic.

\section{Related Work}\label{sec:related-work}

\begin{sloppypar}
\ourparagraph{Automation}
Prior work such as FlowDroid~\cite{arzt_flowdroid_2014}, TaintDroid~\cite{enck_taintdroid_2010,enck_taintdroid_2014}, and AndroidLeaks~\cite{gibler_androidleaks_2012} has focused on how to analyze data flows from apps using static analysis.
In comparison, we capture data further away from the device on the network layer, and focus on capturing real measurements rather than developing scalable methodologies for data collection.
\end{sloppypar}

ReCon~\cite{ren_recon_2016} is another automated approach that instead uses machine learning to detect personal data in network traffic.
To detect personal data, ReCon requires unencrypted network traffic and classifies text as personal data in the categories device identifiers, user identifiers, contact information, location and credentials.
In contrast, the goal of ReCon is to automate the detection of personal data in network traffic, whereas we use a manual keyword-based approach.
Our scopes are also different, as we focus both on the collection and the analysis, and ReCon focuses on the analysis.

Another automated tool is Lumen~\cite{razaghpanah_apps_2018}, which captures and analyzes Android network traffic.
In terms of goals it is one of the works most similar to ours as it includes a measurement study, but its authors also intend for their methodology to be reusable through the tool.
Due to changes in Android since the paper's release in 2018, it is now increasingly difficult to decrypt network traffic, on which Lumen depends to analyze the payloads, which is why we instead depend on mitmproxy to capture and decrypt traffic.

\ourparagraph{Consent}
Prior work has investigated apps' and websites' consent banners.
Kollnig~et~al.~\cite{kollnig_fait_2021} empirically found that free mobile apps most often use third-party trackers, but that only 9.9\% of the 1\,201 analyzed apps obtained consent before starting to collect the data.
Similarly, Paci~et ~al.~\cite{paci_comprehensive_2023} investigated consent banners by studying 400 mobile apps (200 Android apps and 200 iOS apps)
and found that 47\% of the investigated Android apps contacted third-party trackers even when users did not consent to be tracked. 
In the same vein, Morel~et~al.~\cite{morel_tcf_2026} reported that 55.3\% of their investigated Android apps collected data before getting consent.
These works are similar to ours, as all aim to investigate and analyze the data flow from apps.
The main difference between our work and prior works is in methodology and scale: whereas we interact with the apps manually to force ads to trigger, Paci~et~al.~run the apps for a few minutes and Kollnig~et~al.~run the apps for 15 seconds, and Morel~et~al.~solely focus on consent banners.
As such, our study with a smaller sample size (50 apps) and a more targeted category (free-to-play puzzle games with ads) allowed us to manually verify that ads were indeed triggered.
Furthermore, when investigating consent banners, we manually analyzed three specific apps.

\section{Conclusion and Future Work}\label{sec:conclusion}

In this paper, we address the lack of transparency in mobile applications by investigating the network traffic of 50 free-to-play Android games.
We answer the question \emph{What data is collected, and where does it go?} by capturing and analyzing the network traffic with respect to contacted domains, hosts, and endpoints, as well as the actual content of request payloads. 
The analysis of our dataset reveals a complex advertising landscape with many parties involved.
Overall, 239 unique domains are contacted, dominated by ad networks and mediation platforms such as AppLovin, Unity, and DoubleClick. 
Within encrypted HTTPS request payloads we discover information such as device identifiers, timezone and battery information being sent.

Our evaluation of user-side countermeasures shows that using flight mode, an ad-blocking DNS server, or an ad-blocking service like AdGuard can substantially decrease advertising-related requests. 
However, refusing consent and opting out of legitimate interest fail to reliably suppress outbound traffic and can instead introduce inconsistent behavior, such as shifted network destinations or more traffic with the least permissive consent choices.
This observed behavior calls for a deeper analysis into whether apps respect users' privacy choices, and if the apps in fact comply with legal obligations.

Our findings should be interpreted within the limits of our methodology: the number and selection of apps target only one category to prioritize analytical depth over broad generalization. 
The string-matching-based payload search is also sensitive to and biased toward the selection of keywords and regular expressions, which is why we provide open access to our dataset for further analysis. 

\section*{Acknowledgements}
\begin{sloppypar}
 Boel Nelson was partly funded by the Novo Nordisk Foundation, grant NNF25OC0105423. 
\end{sloppypar}

\section*{Disclosure of Interests}
The authors have no competing interests to declare that are relevant to the content of this article.

%
%
%

\bibliographystyle{splncs04}
\bibliography{refs}

\iftoggle{fullpaper}{
\appendix
\renewcommand{\theHsection}{Appendix.\Alph{section}}

\section{Included apps}\label{app:app-list}
This appendix contains additional details about the applications examined in this study. \Cref{tab:apps_examined} list the applications that were included in the measurements, together with their package names, developers, observed advertisement types, emulator availability, and recorded advertisement types.

\noindent Some applications from the initial selection are not included in these tables because they could not be used for the measurements. These excluded applications are listed below, together with the reason for their exclusion. 
Table~\ref{tab:excluded_not_available_emulators} lists the applications that were excluded because they were not available on the emulators used in our experiments.
Table~\ref{tab:excluded_no_ads} lists the applications in which we did not observe any advertisements. 
Finally, Table~\ref{tab:excluded_ad_trigger_too_long} lists the applications we excluded because triggering an advertisement took too long. Since the recordings were performed manually, we set a hard limit of five minutes for triggering the first advertisement in order to keep the measurements consistent and manageable.

{
\begin{table}[htb]
    \rowcolors{2}{white}{\rowstripegray}
    \caption{Applications excluded for not being available on the emulators.}
    \label{tab:excluded_not_available_emulators}
    \centering
    \scriptsize
    \begin{tabularx}{0.95\textwidth}{rXXl}
    \toprule
    \textbf{Rank} & \textbf{Application} & \textbf{Package} & \textbf{Developer} \\
    \midrule

    17 & Magic Sort! & com.grandgames.magicsort & Grand Games A. Ş. \\
    18 & Find Hidden Objects - Spot It! & com.yolo.hiddenobjects & Yolo Game Studios \\
    30 & Match Factory! & net.peakgames.match & Peak \\
    64 & Hexa Diamonds & \makecell[l]{com.remivision.\\HexaDiamonds} & Remi Vision \\
    66 & Bubble Shooter - Classic Pop & bubbleshooter.orig & Bubble Shooter \\
    \bottomrule
    \end{tabularx}
\end{table}
}

{
\begin{table}[htb]
    \rowcolors{2}{white}{\rowstripegray}
    \centering
    \caption{Applications excluded because no advertisements were observed.}
    \label{tab:excluded_no_ads}
    \scriptsize
    \begin{tabularx}{0.95\textwidth}{rXXl}
    \toprule
    \textbf{Rank} & \textbf{Application} & \textbf{Package} & \textbf{Developer} \\
    \midrule

    4 & Royal Kingdom
      & \makecell[l]{com.dreamgames.\\royalkingdom}
      & Dream Games, Ltd. \\
    
    6 & Gossip Harbor®: Merge \& Story
      & \makecell[l]{com.mergegames.\\gossipharbor}
      & Microfun Limited \\
    
    12 & Arrows - Puzzle Escape
       & com.ecffri.arrows
       & Lessmore UG \\
    
    33 & Royal Match
       & \makecell[l]{com.dreamgames.\\royalmatch}
       & Dream Games, Ltd. \\
    
    34 & Yarn Loop
       & com.combo.yarnflow
       & Combo Game \\
    
    37 & Mystery Town: Merge Games
       & \makecell[l]{com.cedargames.\\mysterytown}
       & Cedar Games Studio \\
    
    48 & Color Blast: Block Shooter
       & com.game.block.shooter
       & XGame HK \\
    
    50 & Travel Town - Merge Adventure
       & io.randomco.travel
       & Magmatic Games LTD \\
    
    51 & Clockmaker: Jewel Match 3 Game
       & \makecell[l]{com.belkatechnologies.\\clockmaker}
       & BELKA GAMES \\
    
    56 & Triple Match City
       & com.rc.cityCleaner
       & Ferah Games \\
    
    63 & Merge Restyle
       & com.play.merge.restyle
       & Higgs Gaming \\
    \bottomrule
    \end{tabularx}
\end{table}
}

{
\begin{table}[htb]
    \rowcolors{2}{white}{\rowstripegray}
    \centering
    \caption{Applications excluded as triggering an advertisement took too long.}
    \label{tab:excluded_ad_trigger_too_long}
    \scriptsize
    \begin{tabularx}{0.95\textwidth}{rXXl}
    
    \toprule
    \textbf{Rank} & \textbf{Application} & \textbf{Package} & \textbf{Developer} \\
    \midrule

    5 & Block Crush - Wood Block Blast
      & \makecell[l]{com.wood.block.sudoku.\\puzzle.bm}
      & Wonderful Studio \\
    
    41 & Brilliant Sort: Diamond Puzzle
       & com.belkagames.ds
       & BELKA GAMES \\
       \bottomrule
\end{tabularx}
\end{table}
}

\newcolumntype{C}[1]{>{\centering\arraybackslash}p{\dimexpr#1\textwidth-2\tabcolsep\relax}}
\newcolumntype{L}[1]{>{\raggedright\arraybackslash}p{\dimexpr#1\textwidth-2\tabcolsep\relax}}
\newcolumntype{R}[1]{>{\raggedleft\arraybackslash}p{\dimexpr#1\textwidth-2\tabcolsep\relax}}

{
\rowcolors{2}{white}{\rowstripegray}
\scriptsize
\setlength{\LTcapwidth}{\textwidth}
\begin{longtable}{R{0.10} L{0.26} L{0.18} C{0.145} C{0.13} L{0.185}}
\caption{Overview of the selected applications and observed advertisement types.}
\label{tab:apps_examined}\\
\toprule
\hiderowcolors
\textbf{Rank} & \textbf{Application} & \makecell[l]{\textbf{Observed}\\\textbf{ad types}} & \makecell{\textbf{Accepted}\\\textbf{T\&C /}\\\textbf{Privacy}\\\textbf{Policy}} & \makecell{\textbf{Consent}\\\textbf{for data}\\\textbf{usage}} & \makecell[l]{\textbf{Recorded}\\\textbf{ad types}} \\
\midrule
\showrowcolors
\endfirsthead
\toprule
\hiderowcolors
\textbf{Rank} & \textbf{Application} & \makecell[l]{\textbf{Observed}\\\textbf{ad types}} & \makecell{\textbf{Accepted}\\\textbf{T\&C /}\\\textbf{Privacy}\\\textbf{Policy}} & \makecell{\textbf{Consent}\\\textbf{for data}\\\textbf{usage}} & \makecell[l]{\textbf{Recorded}\\\textbf{ad types}} \\
\midrule
\showrowcolors
\endhead
\midrule
\multicolumn{6}{r}{\textit{Continued on next page}} \\
\endfoot
\bottomrule
\endlastfoot
1 & Block Blast! & rewarded & X & X & rewarded \\
2 & Jigsawcard Solitaire Puzzle & rewarded & X & X & rewarded \\
3 & Pixel Flow! & rewarded & -- & X & -- \\
7 & Hexa Away & banner, rewarded & X & X & banner \\
8 & Block Loop Shooter & rewarded & -- & X & rewarded \\
9 & Tile Explorer - Triple Match & rewarded & X & X & -- \\
10 & Pocket Sort: Coin Merge Puzzle & rewarded, banner & -- & X & -- \\
11 & Differences - Find \& Spot It & banner, rewarded & -- & X & -- \\
13 & Rope Escape Master & rewarded, banner & -- & -- & rewarded, banner \\
14 & Arrow Out & rewarded & -- & X & rewarded, banner \\
15 & Drop Escape: Color Puzzle & rewarded & X & -- & rewarded \\
16 & Jig\-saw\-scapes® - Jigsaw Puzzles & rewarded, interstitial & X & X & rewarded, interstitial \\
19 & Sand Loop & rewarded & X & X & rewarded \\
20 & Sudoku - Classic Sudoku Puzzle & rewarded, interstitial, banner & X & X & rewarded, interstitial, banner \\
21 & Color Block: Combo Blast & banner & -- & -- & banner \\
22 & Goods Puzzle: Sort Challenge™ & rewarded & -- & X & rewarded \\
23 & Tile Clear - Puzzle Game & rewarded & X & X & rewarded \\
24 & Water Out Puzzle™ & rewarded & -- & X & rewarded, interstitial \\
25 & Block Color Mania, Puzzle Game & banner & -- & X & banner \\
26 & Marble Sort! & rewarded & X & X & rewarded \\
27 & Crossmath® - Math Puzzle Games & rewarded, banner, interstitial & -- & X & rewarded \\
28 & Wool Crush™ - Yarn Car Jam & rewarded & -- & X & banner, rewarded \\
29 & Goods Puzzle: 3D Sorting Games & rewarded, banner & -- & X & banner, rewarded \\
31 & Sudoku.com - Classic Sudoku & rewarded, banner & X & X & banner, rewarded, interstitial \\
32 & Jigsolitaire & banner, interstitial & -- & X & banner \\
35 & Prison Puzzle: Brain Twist & banner, interstitial, rewarded & -- & X & rewarded, interstitial \\
36 & Screwdom 3D & banner, rewarded & -- & X & rewarded \\
38 & Sheep Dash: Sheep Away & rewarded & -- & X & rewarded \\
39 & Save The Pets & interstitial, banner, rewarded & -- & X & banner, rewarded, interstitial \\
40 & Color Block Jam & rewarded & X & X & rewarded \\
42 & Number Match - Number Games & banner, rewarded & X & X & rewarded, interstitial \\
43 & Water Sort! Color Sort Puzzle & banner, rewarded & X & X & banner, rewarded \\
44 & Car Out! Traffic Parking Games & banner & X & X & banner \\
45 & Brain Puzzle 3: Crazy Mind & rewarded, banner & -- & X & banner, rewarded \\
46 & Beads Out & rewarded & X & X & rewarded \\
47 & Brain Puzzle 2: Logic Twist & rewarded, banner & -- & X & banner, rewarded \\
49 & Double Tile! & rewarded & X & X & rewarded \\
52 & Block Away - Tap Out Puzzle & banner, rewarded & X & X & banner \\
53 & Queenzle - Queen Puzzle Game & rewarded, banner & -- & -- & banner, rewarded \\
54 & Take Off Bolts: Screw Puzzle & rewarded, banner & X & X & banner, rewarded \\
55 & Gecko Out & rewarded & X & X & banner, rewarded \\
57 & Screw Sort Puzzle: Pin Jam 3D & -- & -- & X & rewarded \\
58 & Connect Master - Match Puzzle & -- & X & X & rewarded \\
59 & Braindom: Brain Games Test & banner, rewarded & X & X & banner \\
60 & Tic Tac Toe 2 Player: XO Game & banner, rewarded & -- & X & rewarded, banner \\
61 & Water SortPuz 3D: Color Sort & rewarded & -- & X & banner was shown under the rewarded ad \\
62 & Fruit Merge™: Match Game & interstitial, rewarded & -- & X & banner, rewarded \\
65 & Brick Blast - Ball Breaker & banner & -- & X & banner \\
67 & DOP 4: Draw One Part & rewarded & -- & X & rewarded, interstitial \\
68 & Arrow Maze - Escape Puzzle & rewarded & -- & X & rewarded \\
\end{longtable}
}
\FloatBarrier

\section{Payload analysis keyword list}\label{app:keyword_list}
This section contains the full keyword list used in the payload analysis. Each table contains one category of keywords.

{
\rowcolors{2}{white}{\rowstripegray}
\begin{longtable}{p{6cm}r}
    \caption{Device specific keywords.}\label{tab:append_device_keywords} \\
    \toprule
    \textbf{Keyword} & \textbf{Description} \\
    \midrule
    \endfirsthead

    \hiderowcolors
    \toprule
    \textbf{Keyword} & \textbf{Description} \\
    \midrule
    \showrowcolors
    \endhead

    \midrule
    \multicolumn{2}{r}{\textit{Continued on next page}} \\
    \endfoot
    \bottomrule
    \endlastfoot

        LTE & value of gsm.network.type \\
        Europe/Belgrade & value of persist.sys.timezone \\
        tablet & value of ro.build.characteristics \\
        no\_require\_sim & part of key: keyguard.no\_require\_sim \\
        UBNX & value of ro.build.host \\
        ro.setupwizard.mode & property name \\
        SM-S9260 & value of ro.product.model \\
        star2qltechn & value of ro.product.device \\
        samsung/star2qltezh/star2qltechn:9/PQ3B. 190801.03250903/G9650ZHU2ARC6:user/release-keys & value of ro.build.fingerprint \\
        star2qltezh-user~9~PQ3B.190801. 03250903 G9650ZHU2ARC6 release-keys & value of ro.build.description \\
        PQ3B.190801.03250903 release-keys & value of ro.build.display.id \\
        ro.debuggable & property name \\
        005274f6 & value of ro.serialno \\
        1774400545 & value of ro.build.date.utc \\
        build.version.extensions.ad\_services & property name \\
        HSPA & value of gsm.network.type \\
        T-Mobile & value of gsm.operator.alpha \\
        persist.radio.airplane\_mode\_on & property name \\
        Europe/Copenhagen & value of persist.sys.timezone \\
        Medium\_Phone\_2 & value of ro.boot.qemu.avd\_name \\
        ro.boot.serialno & property name \\
        EMULATOR36X2X12X0 & value of ro.serialno \\
        google/sdk\_gphone64\_arm64/ emu64a:14/UE1A.230829.036.A4/
        12096271:user/dev-keys & value of ro.bootimage.build.fingerprint \\
        google/sdk\_gphone64\_arm64/emu64a:14/ UE1A.230829.036.A4/12096271:user/release-keys & value of ro.build.fingerprint \\
        sdk\_gphone64\_arm64-user~14U~E1A.230829 .036.A4 12096271 release-keys & value of ro.build.description \\
        emulator & value of ro.build.characteristics \\
        UE1A.230829.036.A4 & value of ro.build.id \\
        emu64a & value of ro.build.product \\
        sdk\_gphone64\_arm64 & value of ro.product.name \\
        qemu & part of property names \\
        ranchu & value of ro.soc.model \\
        arm64-v8a & value of ro.product.cpu.abi \\
        r-d3d21742fc70d910-sb33 & value of ro.build.host \\
        1721146490 & value of ro.build.date.utc \\
        ro.organization\_owned & property name \\
        Pixel 8 Pro & value of ro.product.model \\
        google/husky/husky:14/UP1A.231005.007/ 10750268:user/release-keys & value of ro.product.build.fingerprint \\
        husky & value of ro.build.product \\
        husky-user 14 UP1A.231005.007 10750268 release-keys & value of ro.build.description \\
        gsm.operator.isroaming & property name \\
        gsm.sim.operator.alpha & property name \\
        gsm.operator.iso-country & property name \\
        gsm.sim.operator.iso-country & property name \\
        gsm.operator.alpha & property name \\
        gsm.network.type & property name \\
        net.bt.name & property name \\
        persist.sys.timezone & property name \\
        ro.build.characteristics & property name \\
        ro.com.google.locationfeatures & property name \\
        ro.com.android.dataroaming & property name \\
        build.version.sdk & property name \\
        build.version.release & property name \\
        ro.build.host & property name \\
        en-US & value of locale \\
        ro.build.description & property name \\
        ro.serialno & property name \\
        ro.build.date.utc & property name \\
\end{longtable}
}

{
\small
\rowcolors{2}{white}{\rowstripegray}

\begin{longtable}{p{6cm}>{\raggedleft\arraybackslash}p{6cm}}
    \caption{Keywords from the OpenRTB specification.}
    \label{tab:append_openrtb_keywords} \\

    \toprule
    \textbf{Keyword} & \textbf{Description} \\
    \midrule
    \endfirsthead

    \hiderowcolors
    \toprule
    \textbf{Keyword} & \textbf{Description} \\
    \midrule
    \showrowcolors
    \endhead

    \midrule
    \multicolumn{2}{r}{\textit{Continued on next page}} \\
    \endfoot
    \bottomrule
    \endlastfoot

    rtb & Abbreviation for real-time-bidding \\
    openrtb & Name of standard \\
    BidRequest & Indicates that someone asks for bids \\
    publisher & Entity that controls ad space \\
    SupplyChain & Chain of nodes from beginning to end representing the entities involved in the direct flow of payment for inventory \\
    gdpr & Flag indicating whether the request is subject to GDPR regulations \\
    displaymanager & Name of ad mediation partner, SDK technology, or player responsible for rendering ad \\
    bidfloor & Minimum bid for an ad impression expressed in CPM \\
    BidResponse & Contains bid for a request of ad impression \\
    adid & ID of an ad impression \\

\end{longtable}
}

{
\small
\rowcolors{2}{white}{\rowstripegray}

\begin{longtable}{p{6cm}r}
    \caption{Personal keywords.}
    \label{tab:append_personal_keywords} \\

    \toprule
    \textbf{Keyword} & \textbf{Description} \\
    \midrule
    \endfirsthead

    \hiderowcolors
    \toprule
    \textbf{Keyword} & \textbf{Description} \\
    \midrule
    \showrowcolors
    \endhead

    \bottomrule
    \endfoot

    notevilbob@gmail.com & Email of the used Google account \\
    Bob & Name of the Google account \\
    location & \\
    long & \\
    lat & \\
    country & \\
    language & \\
    ip-address & Regex to match IP addresses \\
    battery & Battery level \\

\end{longtable}
}

{
\small
\rowcolors{2}{white}{\rowstripegray}

\begin{longtable}{p{5.6cm}>{\raggedleft\arraybackslash}p{5.6cm}}
    \caption{Keywords from the list of \cite{towards_user_controlled_privacy}.}
    \label{tab:append_paper_keywords} \\

    \toprule
    \textbf{Keyword} & \textbf{Description} \\
    \midrule
    \endfirsthead

    \hiderowcolors
    \toprule
    \textbf{Keyword} & \textbf{Description} \\
    \midrule
    \showrowcolors
    \endhead
    \midrule
    \multicolumn{2}{r}{\textit{Continued on next page}} \\
    \endfoot
    \bottomrule
    \endlastfoot

    appID & Identifies a specific installation of an app, mainly used by Google libraries \\
    locale & Device language \\
    manufacturer & Device manufacturer \\
    carrier & Information about OS and Mobile Telephone Operator \\
    timezone & Timezone \\
    carrierName & OS \& Mobile Telephone Operator \\
    advertiserId & Advertiser User ID \\
    duration & Event duration \\
    network & Network information \\
    event\_type & Specific event generated by a user \\
    udid & User identifier, used by mopub \\
    device\_id & Device unique identifier \\
    android\_id & ANDROID\_ID \\
    operator & Mobile Telephone Operator \\
    userId & Identifier associated with the device \\
    deviceData & Information about device hardware \\
    appInstanceId & App instance \\
    ad\_id & Google Advertiser ID \\
    installationUuid & Hash associated with an app installed on the device \\
    hardware\_id & ANDROID\_ID \\
    identity & Base64 of UUID and GAID \\
    bssid & MAC address of AP \\
    deviceIP & Device IP \\
    ssid & SSID name \\
    mac\_address & Device MAC address \\
    deviceFingerPrintId & Device fingerprint \\

\end{longtable}
}
\FloatBarrier

\section{Setup runs}\label{app:setup-runs}
This appendix includes the same data and information we reported on the post-setup runs, but for the setup-runs.

\begin{table}[htb]
\centering
    {
    \rowcolors{2}{white}{\rowstripegray}
    \caption{Distribution of request payload content types across all setup runs.}
    \label{tab:request_payload_types_setup_runs}
    \begin{tabularx}{0.6\textwidth}{Xr}
    \toprule
    \textbf{Payload type} & \textbf{Share (\%)} \\
    \midrule
    JSON                    & 63.2 \\
    x-www-form-urlencoded   & 15.9 \\
    Protobuf                & 9.3  \\
    Text                    & 5.1  \\
    Binary                  & 3.6  \\
    Other                   & 3.0  \\
    \bottomrule
    \end{tabularx}
    }
\end{table}

\begin{table}[tb]
    \centering

    \caption{Distribution of search hits by category in all setup runs.}
    \label{tab:payload_hits_category_dist_setup_runs}
    \centering
    {
    \rowcolors{2}{white}{\rowstripegray}
    \begin{tabular}{lrrr}
        \toprule
        \textbf{Category}
        & \textbf{Count}
        & \textbf{Percentage}
        & \textbf{\makecell{Hits/payload\\request}} \\
        \midrule
        Device specific
        & 4\,599 & 32.85 & 0.64 \\
        OpenRTB specification
        & 1\,858 & 13.27 & 0.26 \\
        Personal keywords
        & 2\,909 & 20.28 & 0.41 \\
        Keywords from Caputo~et~al.~\cite{towards_user_controlled_privacy}
        & 4\,632 & 33.09 & 0.65 \\
        \bottomrule
    \end{tabular}
    }
\end{table}

\begin{table}[htb]  
    \caption{Top five domains by search hits in all setup runs.}
    \label{tab:payload_top_domains_setup_runs}
    \centering
    {
    \rowcolors{2}{white}{\rowstripegray}
    \begin{tabular}{lrr}
        \toprule
        \textbf{Domain} & \textbf{Count} & \textbf{Percentage} \\
        \midrule
        unity3d.com & 2\,475 & 17.68 \\
        vungle.com & 2\,214 & 15.82 \\
        googleapis.com & 1258 & 8.99 \\
        amazon-adsystem.com & 1237 & 8.84 \\
        adjust.com & 974 & 6.96 \\
        \bottomrule
    \end{tabular}
    }
\end{table}

\begin{table}[htb]
    \centering
    {
    \rowcolors{2}{white}{\rowstripegray}
    \caption{Top and bottom five apps by search hits in all post-setup runs.}
    \label{tab:payload_top_apps_setup_runs}
    \begin{tabular}{lrr}
        \toprule
        \textbf{(rank) App} & \textbf{Count} & \textbf{Percentage} \\
        \midrule
        (42) Number Match - Number Games & 594 & 4.24 \\
        (31) Sudoku.com - Classic Sudoku & 587 & 4.19 \\
        (29) Goods Puzzle: 3D Sorting Games  & 557 & 3.98 \\
        (40) Color Block Jam & 512 & 3.66 \\
        (27) Crossmath® - Math Puzzle Games & 487 & 3.48 \\
        \midrule
        (22) Goods Puzzle: Sort Challenge™  & 155 & 1.11 \\
        (49) Double Tile! & 151 & 1.08 \\
        ( 8) Block Loop Shooter  & 136 & 0.97 \\
        (38) Sheep Dash: Sheep Away & 134 & 0.96 \\
        (61) Water SortPuz 3D: Color Sort & 133 & 0.95 \\
        \bottomrule
    \end{tabular}
    }
\end{table}

\rowcolors{1}{white}{\rowstripegray}
\begin{table}[htb]
    \caption{Distribution of search hits by content-types in all setup runs.}
    \centering
    \begin{tabular}{lrr}
        \toprule
        \textbf{Type} & \textbf{Count} & \textbf{Percentage} \\
        application/json & 10\,335 & 73.83 \\
        application/x-www-form-urlencoded & 2\,414 & 17.25 \\
        application/x-protobuf & 737 & 5.27 \\
        text/plain & 270 & 1.93 \\
        text/html & 103 & 0.74 \\
        application/x-gzip & 68 & 0.49 \\
        application/protobuf & 33 & 0.24 \\
        application/x-protobuffer & 28 & 0.20 \\
        application/x-sentry-envelope & 9 & 0.06 \\
        application/grpc & 1 & 0.01 \\
        \bottomrule
    \end{tabular}
\end{table}

\FloatBarrier

\section{Additional results}\label{app:additional_results}
{
\rowcolors{1}{white}{\rowstripegray}
\begin{table}[tbh]
    \caption{Distribution of search hits by content-types in all post-setup runs.}\label{tab:payload_hits_content_type_dist_post_setup}
    \centering
    \begin{tabular}{lrr}
        \toprule
        \textbf{Type} & \textbf{Count} & \textbf{Percentage} \\
        application/json & 80\,619 & 73.15 \\
        application/x-www-form-urlencoded & 21\,342 & 19.37 \\
        application/x-protobuf & 5\,019 & 4.55 \\
        text/plain & 1\,502 & 1.36 \\
        text/html & 1\,027 & 0.93 \\
        application/x-gzip & 385 & 0.35 \\
        application/protobuf & 160 & 0.15 \\
        application/x-protobuffer & 66 & 0.06 \\
        application/x-sentry-envelope & 63 & 0.06 \\
        application/grpc & 18 & 0.02 \\
        application/octet-stream & 8 & 0.01 \\
        \bottomrule
    \end{tabular}
\end{table}
}

\iftoggle{morefigures}{
}{
 {
\rowcolors{1}{white}{\rowstripegray}
\begin{table}[tbp]
    \caption{Top and bottom five apps by unique keywords in post-setup runs.}\label{tab:payload_top_apps_keywords_post_setup}
    \centering
    \begin{tabular}{lr}
        \toprule
        \textbf{(rank) App} & \textbf{Unique keywords} \\
        \midrule
        (35) Prison Puzzle: Brain Twist & 38 \\
        (52) Block Away - Tap Out Puzzle & 37 \\
        (24) Water Out Puzzle\texttrademark{} & 37 \\
        (59) Braindom: Brain Games Test & 36 \\
        (67) DOP 4: Draw One Part & 35 \\ \midrule
        (11) Differences - Find \& Spot It & 22 \\
        (38) Sheep Dash: Sheep Away & 22 \\
        (19) Sand Loop & 22 \\
        (43) Water Sort! Color Sort Puzzle & 20 \\
        (68) Arrow Maze - Escape Puzzle & 20 \\
        \bottomrule
    \end{tabular}
\end{table}
}
}

\begin{figure}[tbh]
    \centering
    \includegraphics[width=\linewidth]{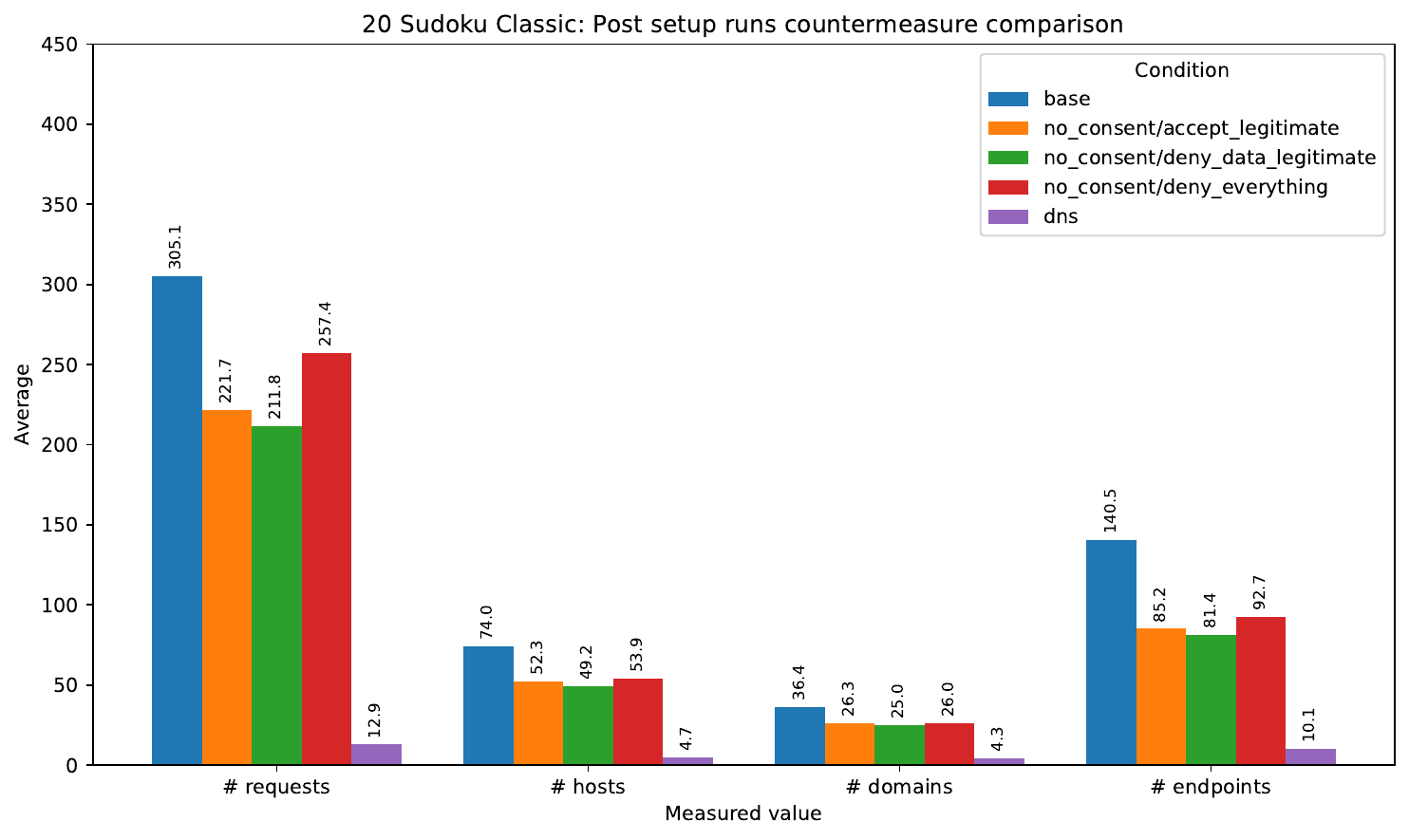}
    \caption{Sudoku: Post-setup runs countermeasure comparison.}
    \label{fig:consent_sudoku_classic_post_setup_runs}
\end{figure}

\begin{figure}[htb]
    \centering
    \includegraphics[width=\linewidth]{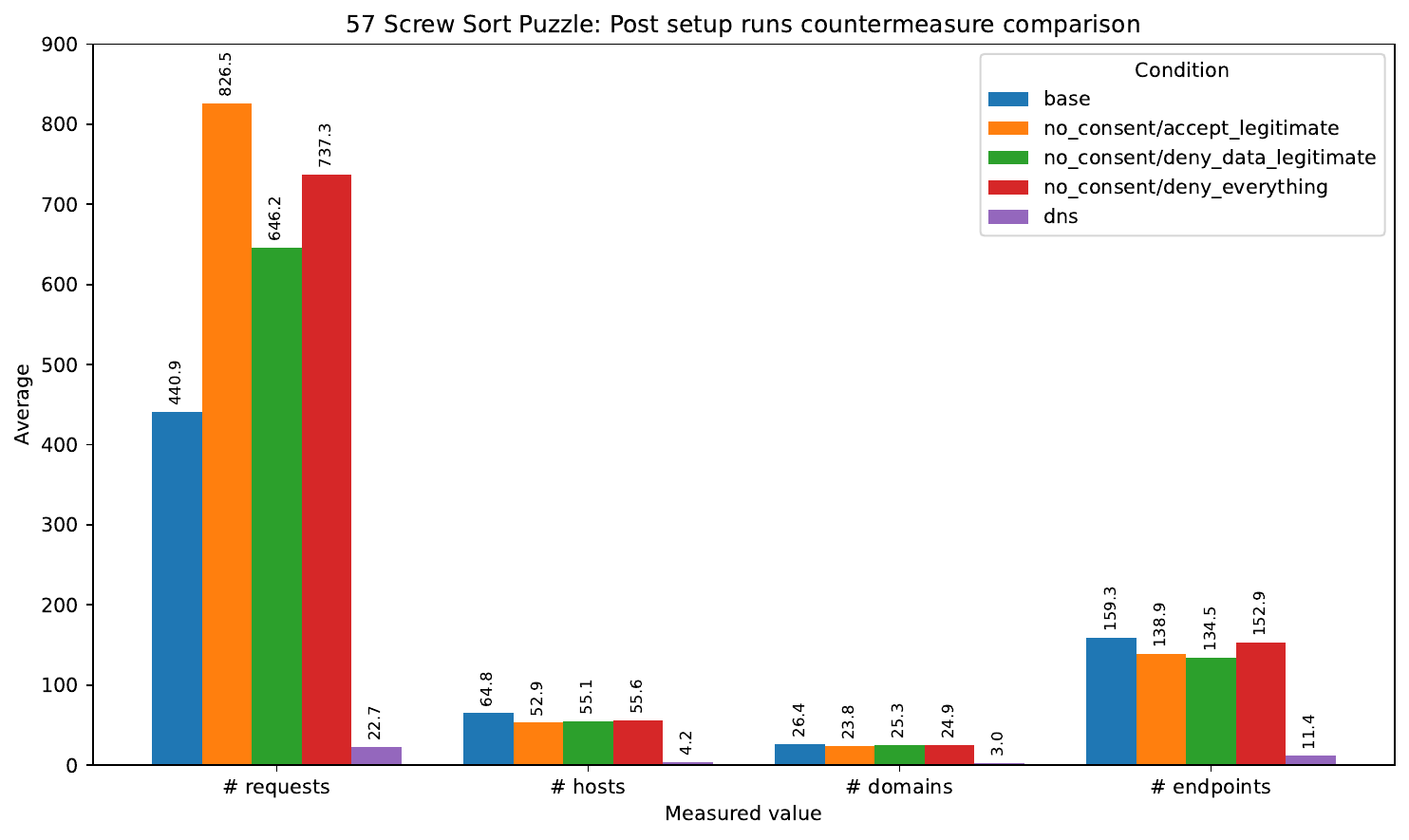}
    \caption{Screw Sort Puzzle: Post-setup runs countermeasure comparison.}
    \label{fig:consent_screw_sort_post_setup_runs}
\end{figure}

\begin{figure}[htb]
    \centering
    \includegraphics[width=\linewidth]{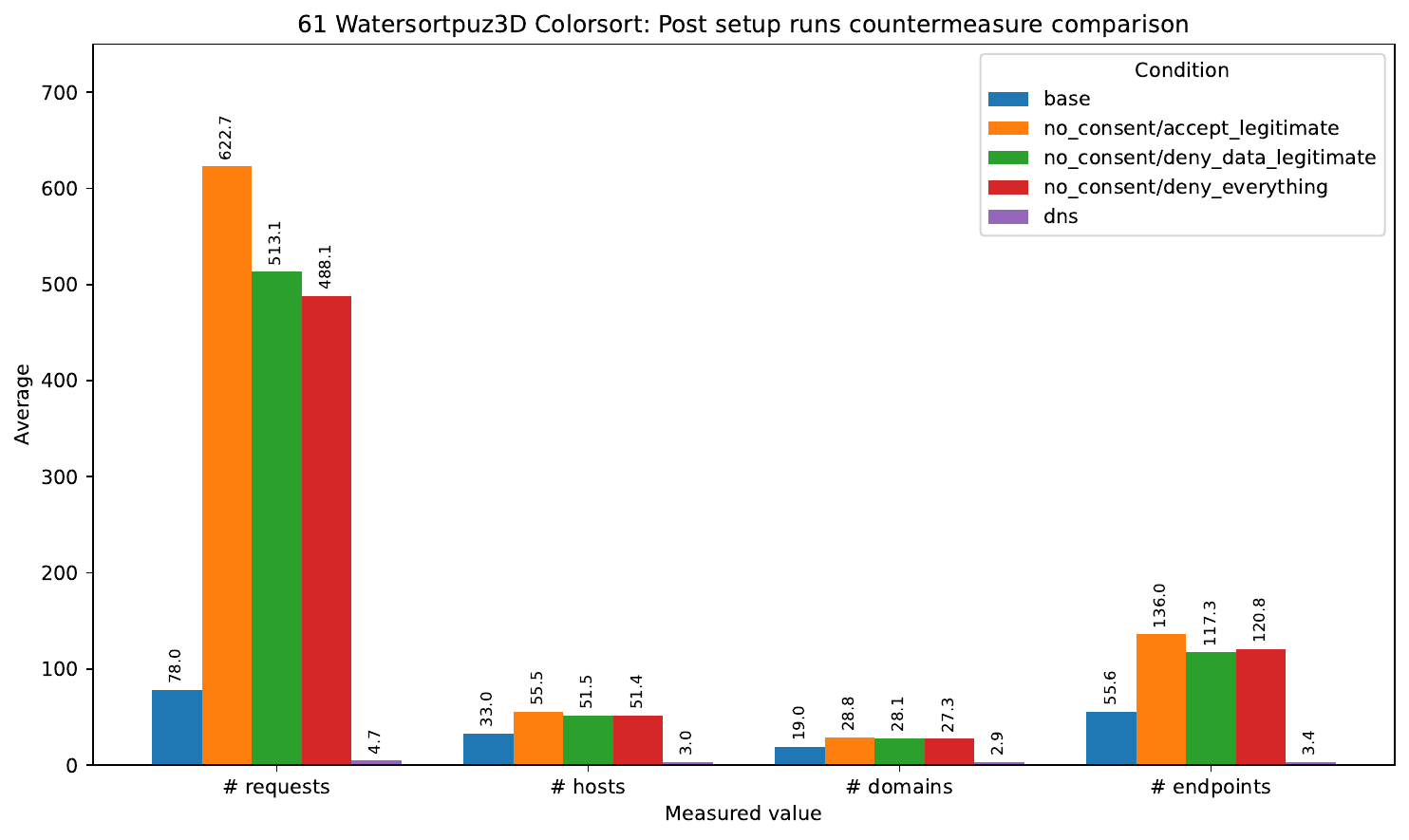}
    \caption{WaterSortPuz 3D: Post-setup runs countermeasure comparison.}
    \label{fig:consent_watersortpuz_post_setup_runs}
\end{figure}

\begin{table}[htb]
    \caption{Top five contacted domains in each post-setup consent-refusal scenario across all selected applications.}
    \label{tab:consent_top_domains}
    \centering
    \begin{tabular}{llrr}
        \toprule\rowcolor{white}
        \textbf{Scenario} & \textbf{Domain} & \textbf{Total} & \textbf{Per run} \\
        \midrule
    
        \rowcolor{\rowstripegray}
        & maticooads.com  & 5\,011 & 167.03 \\
        \rowcolor{\rowstripegray}
        & applovin.com    & 3\,214 & 107.13 \\ 
        \rowcolor{\rowstripegray}
        & gstatic.com     & 1\,428 & 47.60  \\ 
        \rowcolor{\rowstripegray}
        & unity3d.com     & 1\,374 & 45.80  \\ 
        \rowcolor{\rowstripegray}
        \multirow{-5}{*}{\consentaccept}
        & doubleclick.net & 1\,286 & 42.87  \\

        \rowcolor{white}
        & maticooads.com  & 3\,232 & 107.73 \\ 
        & applovin.com    & 2\,822 & 94.07  \\
        \rowcolor{white}
        & gstatic.com     & 1\,512 & 50.40  \\ 
        & unity3d.com     & 1\,030 & 34.33  \\\rowcolor{white}
        \multirow{-5}{*}{\consentdenydata}
        & doubleclick.net & 982    & 32.73  \\

        \rowcolor{\rowstripegray}
        & maticooads.com  & 4\,029 & 134.30 \\ 
        \rowcolor{\rowstripegray}
        & applovin.com    & 3\,014 & 100.47 \\ 
        \rowcolor{\rowstripegray}
        & gstatic.com     & 1\,193 & 39.77  \\ 
        \rowcolor{\rowstripegray}
        & unity3d.com     & 1\,009 & 33.63  \\ 
        \rowcolor{\rowstripegray}
        \multirow{-5}{*}{\consentdenyeverything}
        & fingerhi.net    & 945    & 31.50  \\
        \bottomrule
    \end{tabular}
\end{table}

{
\begin{table}[htb]
\caption{Payload search results for post-setup runs grouped by scenario.}
\label{tab:post-setup-keyword-results-by-scenario}
\centering
    \begin{tabular}{llrrr}
    \toprule
    \textbf{Scenario}
    & \textbf{Category}
    & \textbf{\makecell{Payloads\\with hit}}
    & \textbf{\makecell{Avg. unique\\keywords/payload}}
    & \textbf{Raw hits} \\
    \midrule
    
    \rowcolor{\rowstripegray}
        & Device         & 1\,809 & 1.49 & 3\,643 \\
    \rowcolor{\rowstripegray}
        & OpenRTB        & 940    & 1.10 & 1\,162 \\
    \rowcolor{\rowstripegray}
        & Paper keywords & 1\,938 & 1.52 & 3\,625 \\
    \rowcolor{\rowstripegray}
    \multirow{-4}{*}{\consentbase}
        & Personal       & 1\,293 & 1.43 & 2\,031 \\

    \rowcolor{white}
        & Device         & 6\,100 & 2.70 & 22\,109 \\
        & OpenRTB        & 5\,051 & 1.02 & 5\,252  \\\rowcolor{white}
        & Paper keywords & 6\,137 & 1.80 & 12\,070 \\
    \multirow{-4}{*}{\consentaccept}
        & Personal       & 6\,015 & 1.89 & 11\,686 \\

    \rowcolor{\rowstripegray}
        & Device         & 4\,246 & 2.63 & 15\,096 \\
    \rowcolor{\rowstripegray}
        & OpenRTB        & 3\,249 & 1.03 & 3\,349  \\
    \rowcolor{\rowstripegray}
        & Paper keywords & 4\,195 & 1.80 & 8\,299  \\
    \rowcolor{\rowstripegray}
    \multirow{-4}{*}{\makecell{\consentdenydata}}
        & Personal       & 4\,059 & 1.90 & 8\,052  \\

        \rowcolor{white}
        & Device         & 5\,183 & 2.66 & 18\,591 \\
        & OpenRTB        & 4\,196 & 1.05 & 4\,468  \\\rowcolor{white}
        & Paper keywords & 5\,191 & 1.86 & 10\,716 \\
    \multirow{-4}{*}{\consentdenyeverything}
        & Personal       & 5\,068 & 1.89 & 9\,915  \\
    \bottomrule
    \end{tabular}
\end{table}
}
{
\begin{table}[htb]
    \caption{Per-app keyword matching results for the post-setup runs.}
    \label{tab:per-app-post-setup-keyword-results}
    \centering
    \begin{tabular}{llrrrr}
    \toprule
    \textbf{App}
    & \textbf{Scenario}
    & \textbf{\makecell[l]{Payload\\requests}}
    & \textbf{\makecell[l]{Requests\\with hits}}
    & \textbf{\makecell[l]{Raw\\hits}}
    & \textbf{\makecell[l]{Hits/payload\\request}} \\
    \midrule
    
    \rowcolor{\rowstripegray}
        & \consentbase & 1\,669 & 607 & 4\,033 & 2.42 \\
    \rowcolor{\rowstripegray}
        & \consentaccept & 1\,653 & 720 & 5\,477 & 3.31 \\
    \rowcolor{\rowstripegray}
        & \consentdenydata & 1\,637 & 740 & 5\,618 & 3.43 \\
    \rowcolor{\rowstripegray}
    \multirow{-4}{*}{Sudoku}
        & \consentdenyeverything & 1\,924 & 846 & 6\,180 & 3.21 \\

        \rowcolor{white}
        & \consentbase & 2\,618 & 560 & 2\,541 & 0.97 \\
        & \consentaccept & 7\,399 & 4\,961 & 41\,884 & 5.66 \\\rowcolor{white}
        & \consentdenydata & 5\,447 & 3\,224 & 26\,663 & 4.89 \\
    \multirow{-4}{*}{\makecell[l]{Screw Sort\\Puzzle}}
        & \consentdenyeverything & 6\,498 & 4\,044 & 34\,545 & 5.32 \\

    \rowcolor{\rowstripegray}
        & \consentbase & 2\,452 & 907 & 3\,887 & 1.59 \\
    \rowcolor{\rowstripegray}
        & \consentaccept & 2\,743 & 701 & 3\,756 & 1.37 \\
    \rowcolor{\rowstripegray}
        & \consentdenydata & 2\,208 & 455 & 2\,515 & 1.14 \\
    \rowcolor{\rowstripegray}
    \multirow{-4}{*}{\makecell[l]{Water\\SortPuz 3D}}
        & \consentdenyeverything & 2\,136 & 542 & 2\,965 & 1.39 \\
    \bottomrule
    \end{tabular}
\end{table}
}

\iftoggle{morefigures}{
}{
    {
\begin{table}[htb]
    \caption{Payload search results for post-setup runs grouped by scenario.}
    \label{tab:dns-post-setup-category-payload-summary}
    \centering
    \begin{tabular}{llrrr}

    \toprule
    \textbf{Scenario}
    & \textbf{Category}
    & \textbf{\makecell{Payloads\\with hit}}
    & \textbf{\makecell{Avg. unique\\keywords/payload}}
    & \textbf{Raw hits} \\
    \midrule
    
    \rowcolor{\rowstripegray}
        & Device         & 1\,809 & 1.49 & 3\,643 \\
    \rowcolor{\rowstripegray}
        & OpenRTB        & 940    & 1.10 & 1\,162 \\
    \rowcolor{\rowstripegray}
        & Caputo et al.\cite{towards_user_controlled_privacy} & 1\,938 & 1.52 & 3\,625 \\
    \rowcolor{\rowstripegray}
    \multirow{-4}{*}{Base}
        & Personal       & 1\,293 & 1.43 & 2\,031 \\

        & Device         & 94 & 1.49 & 156 \\
        & OpenRTB        & 30 & 1.00 & 30  \\
        & Caputo et al.\cite{towards_user_controlled_privacy} & 56 & 2.32 & 134 \\
    \multirow{-4}{*}{DNS}
        & Personal       & 44 & 1.00 & 44  \\
    \bottomrule
    \end{tabular}
\end{table}
}
}

\FloatBarrier

{
\rowcolors{2}{white}{\rowstripegray}
\begin{longtable}{lr|lr}
     \caption{All contacted domains and the count of requests towards them.}\label{tab:append_full_domain_list} \\
    \toprule
    \textbf{Domain} & \textbf{Count} & \textbf{Domain} & \textbf{Count} \\
    \midrule
    \endfirsthead

    \hiderowcolors
    \toprule
    \textbf{Domain} & \textbf{Count} & \textbf{Domain} & \textbf{Count} \\
    \midrule
    \showrowcolors
    \endhead

    \midrule
    \multicolumn{4}{r}{\textit{Continued on next page}} \\
    \endfoot
    \bottomrule
    \endlastfoot
        applovin.com & 22684 & freesudoku.me & 23 \\
        unity3d.com & 16953 & presage.io & 22 \\
        doubleclick.net & 10852 & sonicsads.com & 22 \\
        inner-active.mobi & 6418 & unity.cn & 22 \\
        vungle.com & 6350 & lionstudios.cc & 22 \\
        facebook.com & 3849 & nakamacloud.io & 22 \\
        amazon-adsystem.com & 3281 & yahoo.com & 22 \\
        adsmoloco.com & 3276 & funsdata.com & 22 \\
        moloco.com & 3209 & bing.com & 22 \\
        googleapis.com & 3168 & smaato.net & 21 \\
        inmobi.com & 2844 & tenjin.io & 20 \\
        gstatic.com & 2536 & googletagmanager.com & 20 \\
        googlesyndication.com & 2470 & 2mdn.net & 18 \\
        googleusercontent.com & 2396 & cloudfunctions.net & 16 \\
        lunalabs.io & 2238 & bannerflow.net & 16 \\
        appsflyersdk.com & 2209 & smartadserver.com & 15 \\
        afafb.com & 2095 & mindworks-creative.com & 14 \\
        axon.ai & 1700 & maglabmws.com & 14 \\
        tradplusad.com & 1685 & shopifysvc.com & 14 \\
        fingerhi.net & 1515 & wikimedia.org & 13 \\
        liftoff-creatives.io & 1393 & zztfly.com & 12 \\
        adjust.com & 1241 & iposeidoncdn.com & 12 \\
        rollic.gs & 1187 & adtrafficquality.google & 12 \\
        mtgglobals.com & 1111 & mail.ru & 12 \\
        dailyinnovation.biz & 890 & microsoft.com & 12 \\
        saygames.io & 886 & mwsmaglab.com & 12 \\
        easybrain.com & 849 & unpkg.com & 12 \\
        google.com & 733 & clarity.ms & 12 \\
        safedk.com & 720 & cluep.com & 11 \\
        fyber.com & 707 & apple.com & 11 \\
        gameanalytics.com & 699 & icanhazip.com & 11 \\
        bidmachine.io & 688 & 216.227.142.162. & 11 \\
        maticooads.com & 654 & hmgames.org & 11 \\
        yandex.net & 621 & voodoo.io & 11 \\
        ldplayer.net & 557 & adsrvr.org & 11 \\
        suitclub.dk & 512 & tresensa.com & 10 \\
        chartboost.com & 484 & hybrid.ai & 10 \\
        wantfunny.net & 432 & klarna.com & 10 \\
        lazybumblebee.com & 412 & yandexadexchange.net & 9 \\
        googlevideo.com & 396 & dollphoin.site & 9 \\
        kwcdn.com & 332 & inmense.site & 9 \\
        inmobicdn.net & 322 & helpshift.com & 9 \\
        unity3dusercontent.com & 320 & acobt.tech & 9 \\
        temu.com & 314 & facebook.net & 9 \\
        learnings.ai & 291 & mob.com & 8 \\
        klaviyo.com & 287 & amazon.dev & 8 \\
        bucoup.com & 278 & tq-tungsten.com & 8 \\
        liftoff.io & 267 & stackadapt.com & 8 \\
        fbcdn.net & 252 & directservices.it & 8 \\
        castbox.fm & 241 & licdn.com & 8 \\
        nefta.app & 232 & smapps.org & 8 \\
        rayjump.com & 217 & kiwisizing.com & 8 \\
        tiktokpangle.us & 208 & cookieinformation.com & 8 \\
        pubmatic.com & 193 & sentry.io & 7 \\
        privacy-mgmt.com & 170 & simplebro.co & 7 \\
        bobodoo.com & 168 & bravestars.com & 6 \\
        sensorsjourney.com & 167 & google.dk & 6 \\
        fungame.cloud & 166 & 4dvertible.com & 5 \\
        maxesads.com & 142 & zmaticoo.com & 5 \\
        appsflyer.com & 134 & sng.link & 5 \\
        fivecdm.com & 133 & admaster.cc & 5 \\
        sofacompany.com & 118 & linkedin.com & 5 \\
        bytebrew.io & 115 & ytimg.com & 5 \\
        vervegroupinc.net & 110 & changzhi.top & 4 \\
        adjust.net.in & 107 & betano.dk & 4 \\
        doubleverify.com & 102 & iionads.com & 4 \\
        applvn.com & 93 & iion.io & 4 \\
        tiktokpangle-cdn-us.com & 91 & casalemedia.com & 4 \\
        jquery.com & 90 & tpbid.com & 3 \\
        adsafeprotected.com & 88 & mobilefuse.com & 3 \\
        mintegral.com & 82 & cookiepilot.dk & 3 \\
        crashlytics.com & 81 & mouseflow.com & 3 \\
        crushblocks.com & 79 & returnflows.com & 3 \\
        onegg.site & 78 & sleeknote.com & 3 \\
        newsbreak.com & 77 & partnertrackshopify.dk & 3 \\
        cloudfront.net & 77 & trailhub.io & 3 \\
        sensorsdata.cn & 72 & stapecdn.com & 3 \\
        qonversion.io & 69 & shop.app & 3 \\
        pubnative.net & 68 & tiktokw.us & 3 \\
        tripledotapi.com & 68 & profitmetrics.io & 3 \\
        admost.com & 67 & google-analytics.com & 3 \\
        qq.com & 66 & gravatar.com & 3 \\
        dns.google & 66 & wp.com & 3 \\
        ezekielflores.com & 63 & kickoffo.site & 2 \\
        yandex.com & 57 & zendesk.com & 2 \\
        voodoo-tech.io & 57 & yastatic.net & 2 \\
        loomgames.com & 57 & yandex.ru & 2 \\
        usercentrics.eu & 57 & warp90.com & 2 \\
        listdl.com & 56 & ipbase.com & 2 \\
        typekit.net & 56 & bitbucket.io & 2 \\
        googletagservices.com & 53 & webcontentassessor.com & 2 \\
        taxssp.com & 51 & rtb-stack.com & 2 \\
        verve.com & 49 & sascdn.com & 2 \\
        iposeidonbussiness.com & 48 & trustarc.com & 2 \\
        ivymobile.com & 47 & delta-sugar-2024.com & 2 \\
        bm-ads.io & 47 & play-apps.pro & 2 \\
        googleadservices.com & 46 & cylindo.com & 2 \\
        playablefactory.app & 46 & adyen.com & 2 \\
        geoedge.be & 45 & ipify.org & 2 \\
        playfulbytes.games & 44 & g792337340.co & 2 \\
        omjs.com & 43 & everestop.io & 2 \\
        tiktokpangle-b.us & 39 & adjust.io & 2 \\
        shopify.com & 39 & voodoo-adn.com & 1 \\
        amazonaws.com & 36 & js7k.com & 1 \\
        tiktok.com & 35 & yimg.com & 1 \\
        reaktion.com & 34 & gamincatone.com & 1 \\
        zegostudio.com & 33 & sgdn.io & 1 \\
        news-cdn.site & 32 & cloudflareinsights.com & 1 \\
        trustpilot.com & 32 & cloudflare.com & 1 \\
        aliyuncs.com & 30 & azureedge.net & 1 \\
        appmate.io & 30 & xlgmedia.com & 1 \\
        umeng.com & 29 & truste.com & 1 \\
        adnxs.com & 29 & ntesgsght.com & 1 \\
        bugsnag.com & 28 & mookie1.com & 1 \\
        tk0x1.com & 28 & paintdream1010.com & 1 \\
        gvt1.com & 24 & onelink.me & 1 \\
        revenuecat.com & 24 & wexo.io & 1 \\
        a2z.com & 24 & sharethrough.com & 1 \\
        jsdelivr.net & 24 & flashtalking.com & 1 \\
        herodesk.io & 24 \\

\end{longtable}
}

\FloatBarrier

}{

}

\end{document}